\documentclass[%
 preprint,
 amsmath,amssymb,
 aps, physrev,
]{revtex4-2}

\usepackage{graphicx}
\usepackage{dcolumn}
\usepackage{bm}
\usepackage{xcolor}
\usepackage[colorlinks=true, linkcolor=blue, citecolor=red, urlcolor=black]{hyperref}

\newcommand{\rme}{\mathrm{e}}
\newcommand{\rmi}{\mathrm{i}}
\newcommand{\bk}{\boldsymbol{k}}

\definecolor{RED}{named}{red} 

\begin{document}

\preprint{APS/123-QED}

\title{Final states of two-dimensional turbulence above large-scale topography: stationary vortex solutions and barotropic stability}

\author{Jiyang He}
\affiliation{%
 Department of Ocean Science, The Hong Kong University of Science and Technology, Hong Kong, China \\
 Center for Ocean Research in Hong Kong and Macau, The Hong Kong University of Science and Technology, Hong Kong, China.
}
\author{Yan Wang}%
 \email{Contact author: yanwang@ust.hk}
\affiliation{%
 Department of Ocean Science, The Hong Kong University of Science and Technology, Hong Kong, China \\
 Center for Ocean Research in Hong Kong and Macau, The Hong Kong University of Science and Technology, Hong Kong, China.
}%

\date{\today}

\begin{abstract}
The final states of freely decaying two-dimensional (2D) topographic turbulence consist of a background flow and localized vortices. 
While the background flow satisfies a linear potential vorticity (PV)–streamfunction relation, the vortex structures remain poorly understood. 
To address this gap and ensure oceanic relevance, we examine quasi-stationary final states of 2D turbulence over a sinusoidal topography featuring a bump and a dip, where two oppositely signed vortices are locked to the topographic extrema.
After subtracting the background flow, the vortices exhibit a “$\sinh$”-like PV–streamfunction relation, as observed in flat-bottom turbulence. 
Motivated by Gaussian vortex profiles in flat-bottom turbulence,  we propose an empirical model combining the background flow with Gaussian vortices centered at the topographic extrema.
This model accurately reproduces quasi-stationary states and yields locally stationary solutions to the inviscid governing equation.
We further test the model under complex topography and high-energy conditions, confirming that the “$\sinh$”-like trend and Gaussian profiles are robust features of localized vortices.
Linear stability analyses of these stationary vortex solutions reveal background flow-dependent stability: cyclone/elevation and anticyclone/depression configurations are stable at low background energy, while  anticyclone/elevation and cyclone/depression configurations are stable at high background energy. 
These findings align with vortex–topography correlations observed in simulations across energy regimes.
Our results provide explicit vortex solutions for quasi-stationary final states of 2D topographic turbulence and elucidate the mechanism underlying vortex–topography correlations through stability analyses of vortices embedded in topographic background flows.
\end{abstract}


\maketitle

\section{Introduction}
Two-dimensional (2D), quasi-geostrophic (QG) flows over topography are simplified models used to describe large-scale horizontal motions in the atmosphere and ocean, shaped by the presence of mountains, valleys, and variations in seafloor elevation. In atmospheric science, these models are particularly relevant to the study of multiple equilibria associated with blocking patterns over mountainous terrain \cite[see, for example][]{charneyMultipleFlowEquilibria1979}. In oceanography, they provide insight into the formation and stability of quasi-permanent anticyclones trapped in topographic depressions \cite{kohlGenerationStabilityQuasiPermanent2007,solodochFormationAnticyclonesTopographic2021,lacasceVorticesBathymetry2024}, as well as the phenomenon of eddy saturation in the Southern Ocean \cite{constantinouBetaplaneTurbulenceMonoscale2017,constantinouBarotropicModelEddy2018,constantinouEddySaturationSouthern2019}.

The final states toward which two-dimensional (2D) freely decaying topographic turbulence evolves remain poorly understood. In this system, energy undergoes an inverse cascade and is thus approximately conserved, while enstrophy rapidly diminishes due to its forward cascade toward small, dissipative scales. Based on this phenomenology, it has been conjectured that the system minimizes total enstrophy subject to the constraint of conserved total energy \cite{brethertonTwodimensionalTurbulenceTopography1976}. This minimum-enstrophy principle predicts a linear relationship between potential vorticity (hereafter PV) and streamfunction. Solutions to such linear relations yield steady flows aligned with large-scale topography, in which relative vorticity is anti-correlated with topographic elevation.
Although early numerical simulations qualitatively supported the minimum-enstrophy principle \cite{brethertonTwodimensionalTurbulenceTopography1976}, the predicted anti-correlation between flow and topography contradicts oceanic observations of quasi-permanent anticyclones residing over topographic depressions \cite{kohlGenerationStabilityQuasiPermanent2007,solodochFormationAnticyclonesTopographic2021}.

Recent high-resolution direct numerical simulations (DNS) have revealed the general failure of the minimum-enstrophy principle in topographic turbulence \cite{siegelmanTwodimensionalTurbulenceTopography2023}. These simulations show that the final states are dominated by long-lived, isolated vortices embedded within a background flow. The structure of these final states strongly depends on the initial energy: at low energy levels, cyclones and anticyclones are locked to topographic elevations and depressions, respectively, within an inhomogeneous background PV field; at high energy levels, vortices roam freely throughout the domain, accompanied by a nearly homogeneous background PV.
The low-energy solutions are particularly noteworthy, as the emergent vortex-topography correlation aligns with oceanic observations of permanently trapped anticyclones in seafloor bowls \cite{solodochFormationAnticyclonesTopographic2021}.
In our previous work \cite{heMultipleStatesTwodimensional2024}, we further found a strong dependence of the final states on the initial total enstrophy: at low energy levels, higher initial enstrophy leads to final states with more energetic and abundant vortices, along with a more homogenized background PV field; 
at high energy levels, a sufficiently low initial total enstrophy—introduced at the domain scale—can also induce topographic locking of vortices, albeit in an anti-correlated manner.

Localized vortices are commonly observed in the final states of two-dimensional (2D) freely decaying turbulence without topography (hereafter referred to as 2D flat-bottom turbulence) \cite{mcwilliamsEmergenceIsolatedCoherent1984,mcwilliamsVorticesTwodimensionalTurbulence1990}. These final states are typically dominated by a pair of opposite-signed vortices \cite{matthaeusDecayingTwodimensionalNavierStokes1991,matthaeusSelectiveDecayCoherent1991}, a phenomenon that can be interpreted as the clustering of like-signed point vortices, according to predictions from equilibrium statistical mechanics \cite{onsagerStatisticalHydrodynamics1949}.
Moreover, the final states of 2D flat-bottom turbulence are found to satisfy a “$\sinh$” relationship between vorticity and streamfunction \cite{montgomeryRelaxationTwoDimensions1992}, derived from a maximum-entropy principle \cite{joyceNegative1973,MontgomeryStatistical1974}. The shapes of the localized vortices in these flows are typically Gaussian \cite{mcwilliamsVorticesTwodimensionalTurbulence1990,jimenezStructureVorticesFreely1996}.

Building upon insights from 2D flat-bottom turbulence, we aim to examine the final states of 2D topographic turbulence and characterize the localized vortices—specifically, their adherence to the “$\sinh$” vorticity–streamfunction relation and the Gaussian spatial structure. 
These analyses require a suitable decomposition of the flow into localized vortices and a background flow. 
First, we assess the validity of the ``$\sinh$'' relation for localized vortices by isolating them from the topographic background flow. 
Next, we develop and validate an empirical model that represents the final state as a superposition of a background flow and localized vortices. 
The background flow is assumed to follow a linear PV–streamfunction relation, as established in previous studies \cite{siegelmanTwodimensionalTurbulenceTopography2023,heMultipleStatesTwodimensional2024}, while the vortices are modeled as Gaussian structures that follow the ``$\sinh$'' relation. 
We demonstrate that this empirical model performs well for low-energy quasi-stationary states, where vortices are centered precisely at the topographic extrema. 
Thus, the model provides a class of stationary solutions for quasi-stationary final states and, for the first time, explicitly incorporates the structure of localized vortices. 
This quasi-stationary regime is geophysically relevant, as quasi-stationary and quasi-permanent anticyclones are commonly observed in topographically confined ocean basins \cite{solodochFormationAnticyclonesTopographic2021}.

The random multi-scale topography used in previous studies \cite{siegelmanTwodimensionalTurbulenceTopography2023,heMultipleStatesTwodimensional2024} contains numerous local extrema across all spatial scales, allowing for complex spatial distributions of localized vortices with varying sizes and strengths. 
This complexity poses significant challenges for quantitative modeling. 
Since our primary goal is to test the validity of the modeling strategy, we mainly focus on an idealized large-scale sinusoidal topography with only one elevation and one depression, which permits a much simpler vortex distribution.
Similar idealized topographies have been employed in simulations of barotropic and baroclinic QG turbulence \citep{vallis1993generation,thompson2010jet}. 
However, previous studies have not reported the formation of isolated vortices locked to topography, possibly due to factors such as insufficient grid resolution, large-scale initial conditions, flow baroclinicity, or external forcing.

One puzzling phenomenon in 2D topographic turbulence, also observed in the ocean, is the correlation between vortices and topography, where cyclones and anticyclones tend to align with the respective topographic elevations and depressions at low energy levels. 
Previous studies \cite{kohlGenerationStabilityQuasiPermanent2007,siegelmanTwodimensionalTurbulenceTopography2023} have attributed this behavior to the self-propagation of vortices induced by topographic PV gradient, a process termed $\beta$-drift: cyclones tend to migrate upslope toward the center of an elevation, whereas anticyclones descend toward the center of a depression \cite{carnevalePropagationBarotropicVortices1991}. 
The observed vortex-topography correlation can also be interpreted through the lens of vortex stability above topography, as examined via linear stability analyses of exact vortex solutions. 
For example, Ref. \cite{benilovStabilityTwoLayerQuasigeostrophic2005} analyzed circular isolated vortices over circular topography in a two-layer QG system and found that a topographic elevation stabilizes cyclones and destabilizes anticyclones, while a depression has the opposite effect. 
In Ref. \cite{gonzalezLinearStabilityMonopolar2023}, classical stability theorems were generalized to include topographic effects, providing necessary stability conditions for vortices with arbitrary radial profiles. 
Based on vortex solutions derived by the same authors \cite{gonzalezQuasigeostrophicVortexSolutions2021}, they concluded that anticyclone/mountain and cyclone/valley configurations may be unstable, whereas cyclone/mountain and anticyclone/valley configurations are generally stable. 
In this work, we conduct similar linear stability analyses of the stationary vortex solutions we develop. 
These solutions are superpositions of background flows with underlying linear PV–streamfunction relations and Gaussian vortices. 
While the flows with linear PV–streamfunction relations have been shown to be nonlinearly stable \cite{carnevaleNonlinearStabilityStatistical1987,nycanderStableUnstableVortices2004}, we focus on the stability of Gaussian vortices of different polarities embedded within spatially varying background PV fields. Our goal is to uncover vortex stability properties consistent with the vortex–topography correlation and anti-correlation observed at the respective low and high energy levels in our previous simulations \cite{heMultipleStatesTwodimensional2024}.

This paper is organized as follows.
Section~\ref{sec:configurations} introduces the problem configurations, including the governing equations, topography, the exact stationary solutions based on a linear PV–streamfunction relationship, and the numerical set-ups.
Section~\ref{sec:stationary_vortex_solutions} is devoted to the development of stationary vortex solutions.
In \S~\ref{sec:phenomenology}, we describe the phenomenology of the final states at low energy levels observed in numerical simulations.
We then examine the scatter plots of PV versus streamfunction to identify the functional relationships characterizing the localized vortices in \S~\ref{sec:q-psi_relation}.
Thereafter, we construct an empirical model for the quasi-stationary final states and evaluate its performance against the simulation results in \S~\ref{sec:empirical_model}.
Finally, we discuss the applicability of our framework to final states above complex random topography and at high energy levels in \S~\ref{sec:random_topography_and_high_energy}.
In \S~\ref{sec:stability}, we conduct linear stability analyses of our stationary vortex solutions on a circular disk around a topographic extremum.
The conclusion is given in \S~\ref{sec:conclusion}.

\section{Configurations}
\label{sec:configurations}

\subsection{Governing equations}

We consider 2D topographic turbulence described by an unforced, single-layer QG flow on an $f$-plane above topography in a $[-\pi,\pi)\times [-\pi,\pi)$ doubly periodic domain, thus with the domain length defined by $L=2\pi$.
The governing equation of the QG model reads
\begin{equation}\label{eq:qg_equation}
    \frac{\partial q}{\partial t} + J(\psi, q) = D\zeta,
\end{equation}
where $q(x,y,t)$ denotes the QG PV, $\psi(x,y,t)$ represents the geostrophic streamfunction, $\zeta(x,y,t)$ stands for the relative vorticity,
$J(\psi, q) = \partial\psi/\partial x\cdot\partial q/\partial y-\partial\psi/\partial y\cdot\partial q/\partial x$ is the Jacobian, and
$D\zeta$ represents the dissipation term to be specified in the subsequent numerical section.
The zonal and meridional velocities are obtained as $u=-\partial\psi/\partial y$ and $v=\partial\psi/\partial x$, respectively.

The QG PV is related to the geostrophic streamfunction through
\begin{equation}\label{eq:q_psi_differential_relation}
    \quad q=\zeta + \eta = \Delta\psi + \eta.
\end{equation}
Here, $\eta(x,y)=-f_0h_1/h_0$ denotes the topographic PV for a total depth of $h_0+h_1(x,y)$ with small fluctuations $h_1(x,y)$ around a constant average depth $h_0$, scaled by the local Coriolis frequency $f_0$;
$\Delta$ is the 2D Laplacian operator.
In the absence of topography, the QG PV equation (\ref{eq:qg_equation}) reduces to that for 2D flat-bottom turbulence.

\begin{figure}
    \centering
    \includegraphics[width=1\linewidth]{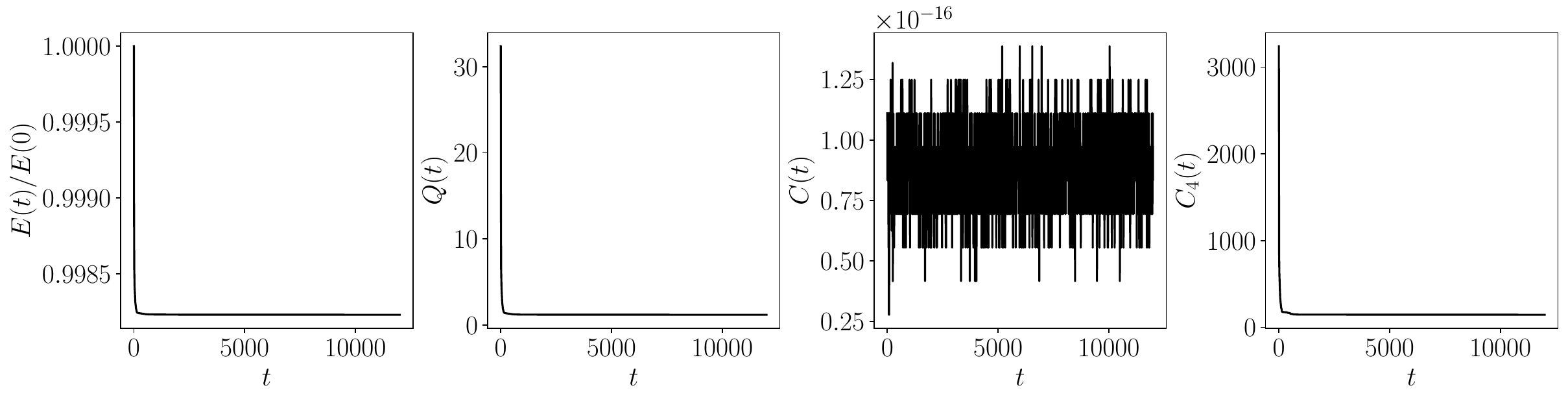}
    \caption{Time series of integrals of a typical simulation of 2D topographic turbulence with zero initial total circulation: $(a)$ energy $E(t)$ scaled by initial value; 
    $(b)$ total enstrophy $Q(t)$;
    $(c)$ total circulation $C(t)$;
    $(d)$ quartic Casimir $C_4(t)$.}
    \label{fig:integral_time_series}
\end{figure}

In the inviscid limit ($D\zeta=0$), the QG PV equation (\ref{eq:qg_equation}) admits an infinite number of integral invariants,
including the energy
\begin{equation}\label{eq:energy}
    E = \frac{1}{2L^2}\int_{-\pi}^{\pi}\int_{-\pi}^{\pi} |\nabla\psi|^2 dxdy
\end{equation}
and a series of Casimirs
\begin{equation}\label{eq:casimir}
    C_n = \frac{1}{nL^2}\int_{-\pi}^{\pi}\int_{-\pi}^{\pi} q^n dxdy.
\end{equation}
For the latter, $n=1$ and $n=2$ correspond to the total circulation (denoted as $C$) and the total enstrophy (denoted as $Q$), respectively.
As shown in Fig.\ref{fig:integral_time_series}, in numerical simulations of decaying turbulence, the energy $E$ decreases slowly and can be regarded as being nearly conserved; the total enstrophy decreases drastically.
For non-quadratic invariants, the first-order Casimir (total circulation, $C$) does not change over time, while the  quartic Casimir $C_4$ decreases substantially as the total enstrophy $Q$ does.

\subsection{Topography}

\begin{figure}
    \centering
    \includegraphics[width=0.5\linewidth]{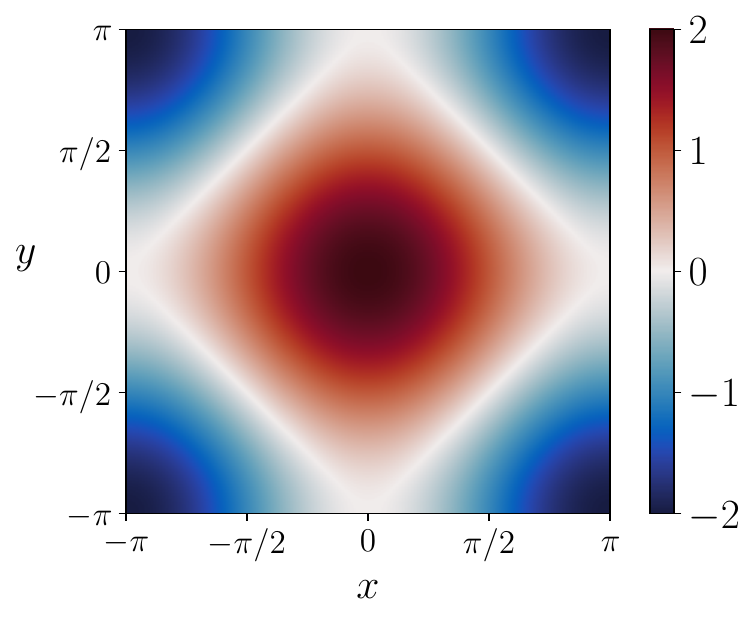}
    \caption{Topography: $\eta(x,y)=\cos{x}+\cos{y}$.}
    \label{fig:topography}
\end{figure}

Throughout this work, we consider an idealized large-scale topography that is a sinusoidal bump plus a dip,
\begin{equation}\label{eq:topography}
    \eta(x,y)=\cos{x}+\cos{y},
\end{equation}
as shown in Fig. \ref{fig:topography}.
In this configuration, there are one maximum and one minimum of topographic PV.
The topographic maximum and minimum are located at the domain center ($x=0,y=0$) and the four corners ($x=\pm\pi, y=\pm\pi$), respectively, due to horizontal periodicity.
This topography allows for a very simple spatial arrangement of localized vortices, with only two vortices positioned above the respective topographic extrema. 
Such a configuration offers a convenient framework for examining vortex structures and performing quantitative modeling.

\subsection{Linear PV-streamfunction relation}
The classical minimum-enstrophy principle yields an exact solution to the Jacobian $J(\psi,q)=0$.
It is characterized by a linear PV-streamfunction relation \cite{brethertonTwodimensionalTurbulenceTopography1976},
\begin{equation}\label{eq:linear_relation}
    q_* = \Delta\psi_* + \eta =\mu\psi_*,
\end{equation}
where $\mu$ is a linear factor.
Given a topography $\eta(x,y)$, the stationary relative vorticity $\zeta_*(x,y,\mu)$, PV $q_*(x,y,\mu)$ and streamfunction $\psi_*(x,y,\mu)$ can be readily obtained by solving the above equation.
For the topography shown in Fig. \ref{fig:topography}, the results are 
\begin{subequations}\label{eq:minimum_enstrophy_solution}
    \begin{eqnarray}
        \zeta_*(x,y,\mu) &=& -\frac{1}{\mu+1}(\cos{x} + \cos{y}), \\
        q_*(x,y,\mu) &=& \frac{\mu}{\mu+1}(\cos{x} + \cos{y}), \\
        \psi_*(x,y,\mu) &=& \frac{1}{\mu+1}(\cos{x} + \cos{y}).
    \end{eqnarray}
\end{subequations}
The energy and total enstrophy are found to satisfy
\refstepcounter{equation}\label{eq:minimumenstrophy_energy_entrophy}
$$
E_*(\mu)=\frac{1}{2(\mu+1)^2}; \quad Q_*(\mu)=\frac{\mu^2}{2(\mu+1)^2}, \eqno{(\theequation{\text{a},\text{b}})}
$$
respectively.
In the above formulae, $\mu=-1$ represents a singular point, separating two branches of solutions with $\mu>-1$ and $\mu<-1$. 
In line with Ref.~\cite{siegelmanTwodimensionalTurbulenceTopography2023,heMultipleStatesTwodimensional2024}, we restrict $\mu$ to the former range, $(-1,+\infty)$.

Although a simple linear relation cannot fully describe the final states of 2D topographic turbulence in most cases, it still carries predicebative skills. On the one hand, it effectively characterizes the background topographic flow once localized vortices are removed \cite{siegelmanTwodimensionalTurbulenceTopography2023,heMultipleStatesTwodimensional2024}. On the other hand, it defines a critical case where the linear factor $\mu = 0$. With $\mu = 0$, the PV is homogenized ($q_\# = 0$), and the corresponding streamfunction $\psi_\#$—referred to as the topographic streamfunction \cite{siegelmanTwodimensionalTurbulenceTopography2023}—is obtained by applying the inverse Laplacian operator to the topography:
\begin{equation}\label{eq:topographic_streamfunction}
    \psi_\# = -\Delta^{-1}\eta.
\end{equation}
The associated energy, denoted as $E_\#=E_*(\mu=0)$, has been used as a threshold between low- and high-energy regimes, and for setting initial energy levels in simulations \cite{siegelmanTwodimensionalTurbulenceTopography2023,heMultipleStatesTwodimensional2024}.
The low  and high energy levels correspond to positive and negative linear factors $\mu$, respectively, as demonstrated by the energy expression (Eq.\ref{eq:minimumenstrophy_energy_entrophy}a).
At low energy levels ($\mu>0$), the relative vorticity expression (\ref{eq:minimum_enstrophy_solution}a) yields anticyclonic ($\zeta_*<0$) and cyclonic ($\zeta_*>0$) flows above the respective topographic elevation (at the domain center) and depression (at the corners), which means that the flow and topography with an underlying linear PV-streamfunction  are anti-correlated (see Eq. 3 in Ref. \cite{siegelmanTwodimensionalTurbulenceTopography2023}).

\subsection{Numerical set-ups}
We numerically solve the QG PV equation using the open-source, GPU-accelerated pseudo-spectral package \texttt{GeophysicalFlows.jl} \cite{constantinou2021geophysicalflows}. The computational domain is set to $[-\pi, \pi) \times [-\pi, \pi)$, corresponding to a domain size of $L = 2\pi$, with the smallest wavenumber equal to $1$. All simulations are performed at a spatial resolution of $1024 \times 1024$.
Dissipation is implemented via a spectral filter applied to the relative vorticity at high wavenumbers at the end of each time step. Time integration is carried out using a fourth-order Runge–Kutta scheme with a fixed time step of $0.002$, ensuring that the Courant–Friedrichs–Lewy (CFL) number remains below 1. Each simulation is run for $1.5 \times 10^7$ time steps to ensure convergence to a final state for which there is no apparent temporal variation of the total enstrophy (see Fig.\ref{fig:integral_time_series}) and the localized vortices cease to merge.
The flow structures of the final states are then analyzed.

\begin{figure}
    \centering
    \includegraphics[width=1\linewidth]{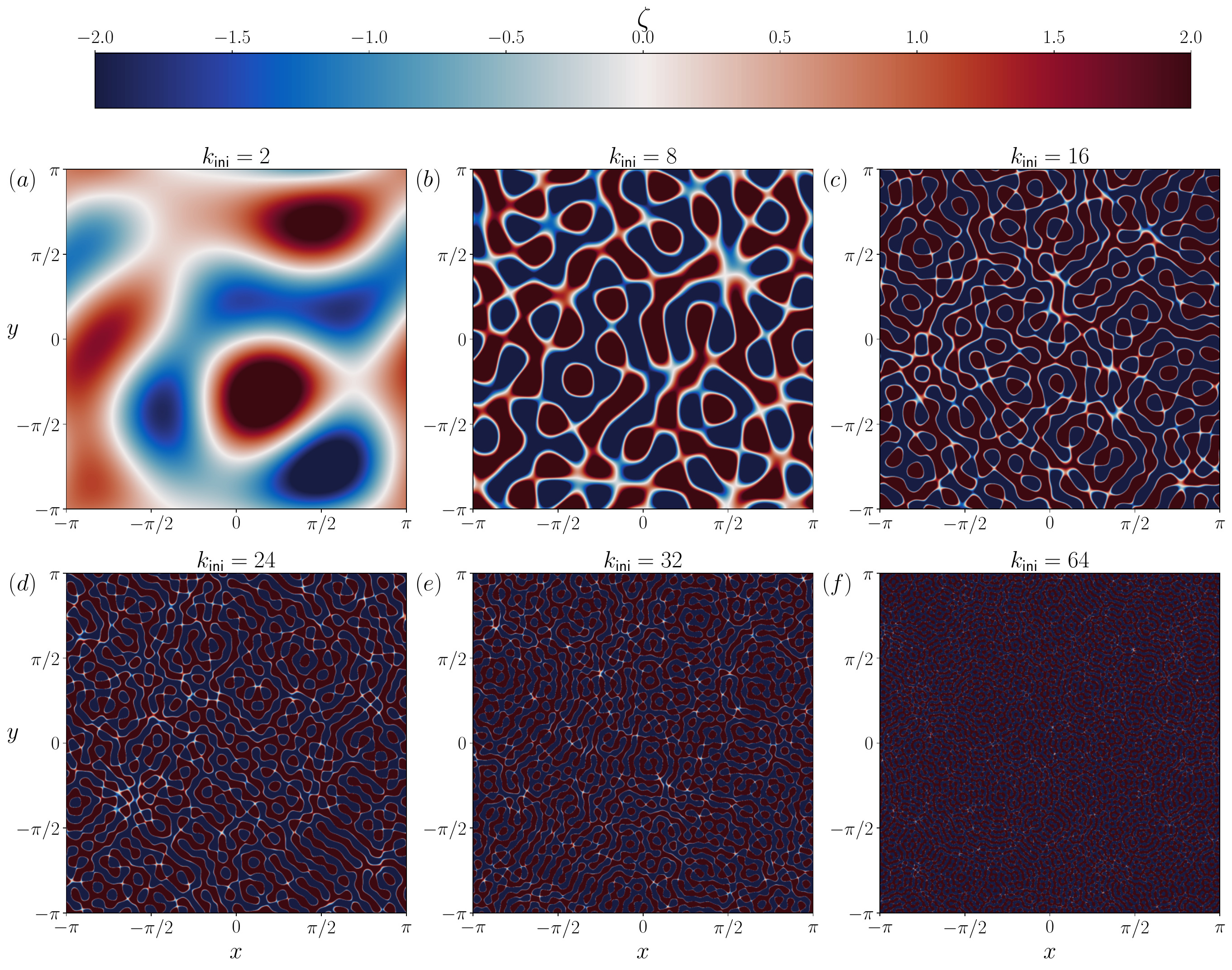}
    \caption{Initial monoscale fields of the relative vorticity $\zeta$ centering around the wavenumbers $k_{ini}=[2,8,16,24,32,64]$ ($a$--$f$), but with the same energy level $0.25E_\#$.}
    \label{fig:initial_conditions}
\end{figure}

To explore the range of possible final states in 2D freely-decaying topographic turbulence, we employ a series of initial conditions following the approach in our previous work \cite{heMultipleStatesTwodimensional2024}. 
Each simulation is initialized with a structureless random field of relative vorticity, $\zeta(x, y, 0)$, concentrated around a specified centroid wavenumber $k_{\text{ini}}$, with spectral content confined to the narrow band $[k_{\text{ini}} - 0.5,, k_{\text{ini}} + 0.5]$. These monoscale initial conditions are characterized by two parameters: the centroid wavenumber $k_{\text{ini}}$ and the total energy $E$.
We are mainly interested in the low energy regimes with quasi-stationary vortices precisely locked to topography, for which stationary vortex solutions can be developed.
The energy of all low-energy simulations is fixed at $E=0.25 E_\#$.
Our next interest lies in how varying the initial total enstrophy—achieved by changing $k_{\text{ini}}$—affects the resulting final states at this fixed energy level. Higher $k_{\text{ini}}$ values correspond to smaller-scale initial structures and thus higher initial enstrophy.
The initial relative vorticity fields for $k_{\text{ini}} = [2, 8, 16, 24, 32, 64]$ are illustrated in Fig.~\ref{fig:initial_conditions}.

\section{Stationary vortex solutions}
\label{sec:stationary_vortex_solutions}

In this section, we aim to develop stationary vortex solutions for the low-energy quasi-stationary final states above the sinusoidal topography (Fig.\ref{fig:topography}).
The results are presented with the description of the phenomenology of the final states, the examination of the relationship between PV and streamfunction and the construction of an empirical model of the stationary vortex solutions.
Finally, we discuss the applicability of the framework to final states above complex random topography and at high energy levels, where the group of vortices exhibit complex spatial distributions and mobility.

\subsection{Phenomenology of final states}
\label{sec:phenomenology}

\begin{figure}
    \centering
    \includegraphics[width=\textwidth]{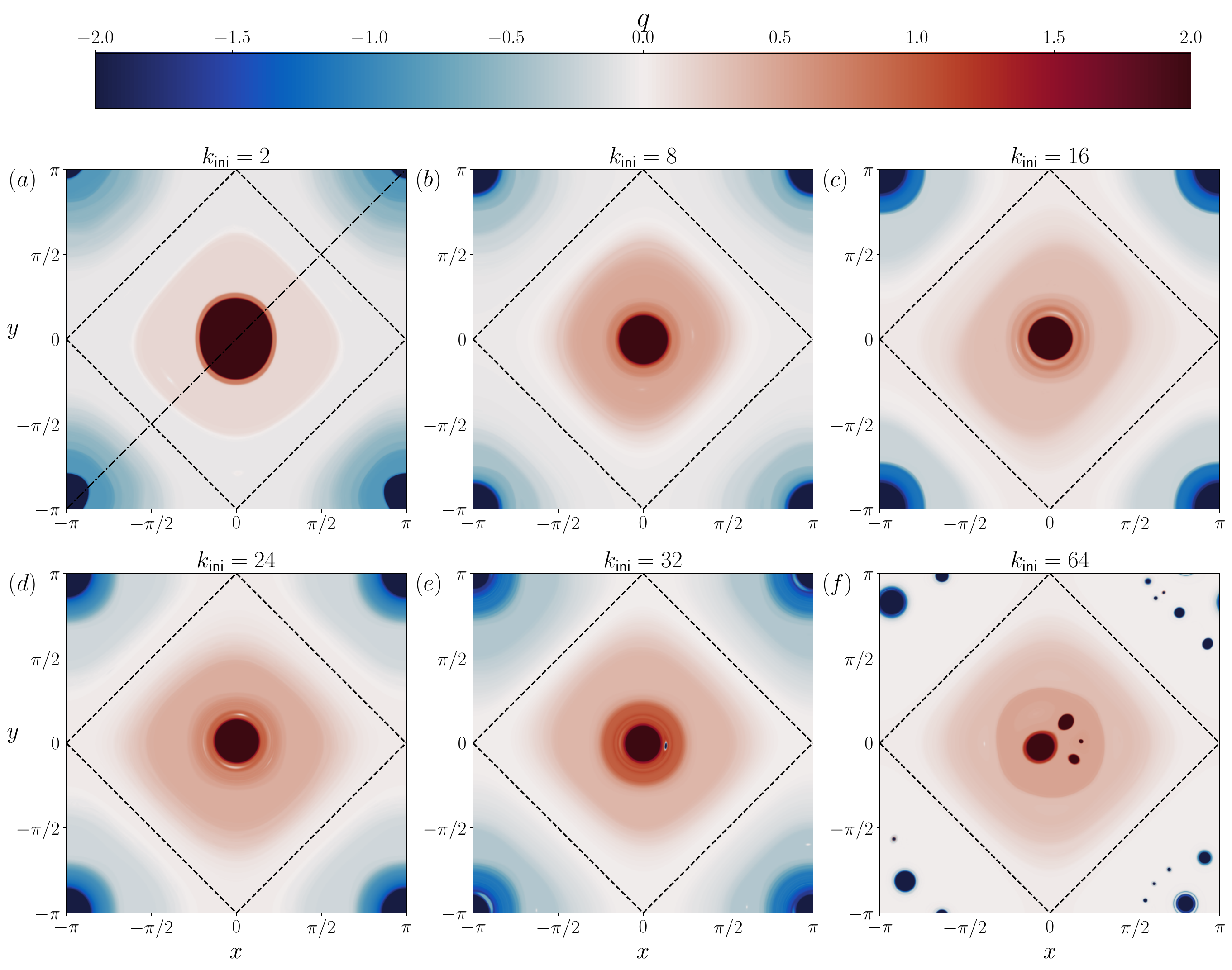}
    \caption{Final states of PV at $t=30000$ evolving from the initial fields (Fig. \ref{fig:initial_conditions}) above the sinusoidal topography (Fig. \ref{fig:topography}).
    Dashed lines denote $\eta=0$ and separate the areas of $\eta>0$ (around the centre) and $\eta<0$ (around the corners).}
    \label{fig:bump_contour}
\end{figure}

Fig.~\ref{fig:bump_contour} shows the final states of potential vorticity (PV) at time $t = 30{,}000$, resulting from the different initial conditions illustrated in Fig.~\ref{fig:initial_conditions}, all simulated over the sinusoidal topography shown in Fig.~\ref{fig:topography}. Short-term evolutions from these final states are provided in the supplementary movies \cite{SupplementalMovie}. Despite the long integration times required to reach these states, noticeable unsteadiness persists.
In all cases, cyclones ($q > 0$) cluster near the topographic maximum (center of the domain), while anticyclones ($q < 0$) gather around the topographic minimum (corners of the domain). However, the characteristics of these vortices—such as their unsteadiness, number, shape, size, and strength—vary significantly with the initial wavenumber.
For the $k_{\text{ini}} = 2$ case (Fig.~\ref{fig:bump_contour}$a$), the resulting two vortices exhibit pronounced unsteadiness (see the corresponding movie \cite{SupplementalMovie}): the cyclone's edge deforms over time, and the anticyclone is not fully locked to the topography, drifting around the topographic minimum.
For intermediate wavenumbers $k_{\text{ini}} \in [8, 16, 24, 32]$ (Fig.~\ref{fig:bump_contour}$b$–$e$), both cyclones and anticyclones are nearly circular and centered precisely over the topographic maximum and minimum, respectively. The corresponding movies \cite{SupplementalMovie} show that these vortices are nearly stationary. 
Notably, a small anticyclone persists near the cyclone in the $k_{\text{ini}} = 32$ case (Fig.~\ref{fig:bump_contour}$e$).
In the $k_{\text{ini}} = 64$ case, multiple cyclones and anticyclones emerge as vortex crystals surrounding the topographic extrema. 
As shown in the movie \cite{SupplementalMovie}, the cyclones undergo solid-body rotation, while the anticyclones remain nearly frozen. 
The asymmetry may be connected to the unequal spatial distributions of the initial condition (Fig.\ref{fig:initial_conditions}$f$) around the topographic maximum and minimum.
The vortex crystals resemble the polar vortex structures observed on a $\gamma$-plane in Ref.\cite{siegelmanPolarVortexCrystals2022}. 
This resemblance is intuitive, as the sinusoidal topography near its extrema varies approximately as $\eta \sim r^2$ (where $r$ is the radial distance to the extremum), thereby inducing an analogous $\gamma$-effect similar to that over a polar cap.

In summary, at low energy levels, the final states at intermediate wavenumbers ($k_{\text{ini}} \in [8, 16, 24, 32]$) consist of two quasi-stationary vortices centered at the respective topographic extrema. 
For these cases, simple functional relationships between PV and streamfunction can be identified, enabling the construction of semi-analytical stationary vortex solutions as follows.

\subsection{PV-streamfunction relations}
\label{sec:q-psi_relation}

\begin{figure}
    \centering
    \includegraphics[width=1\textwidth]{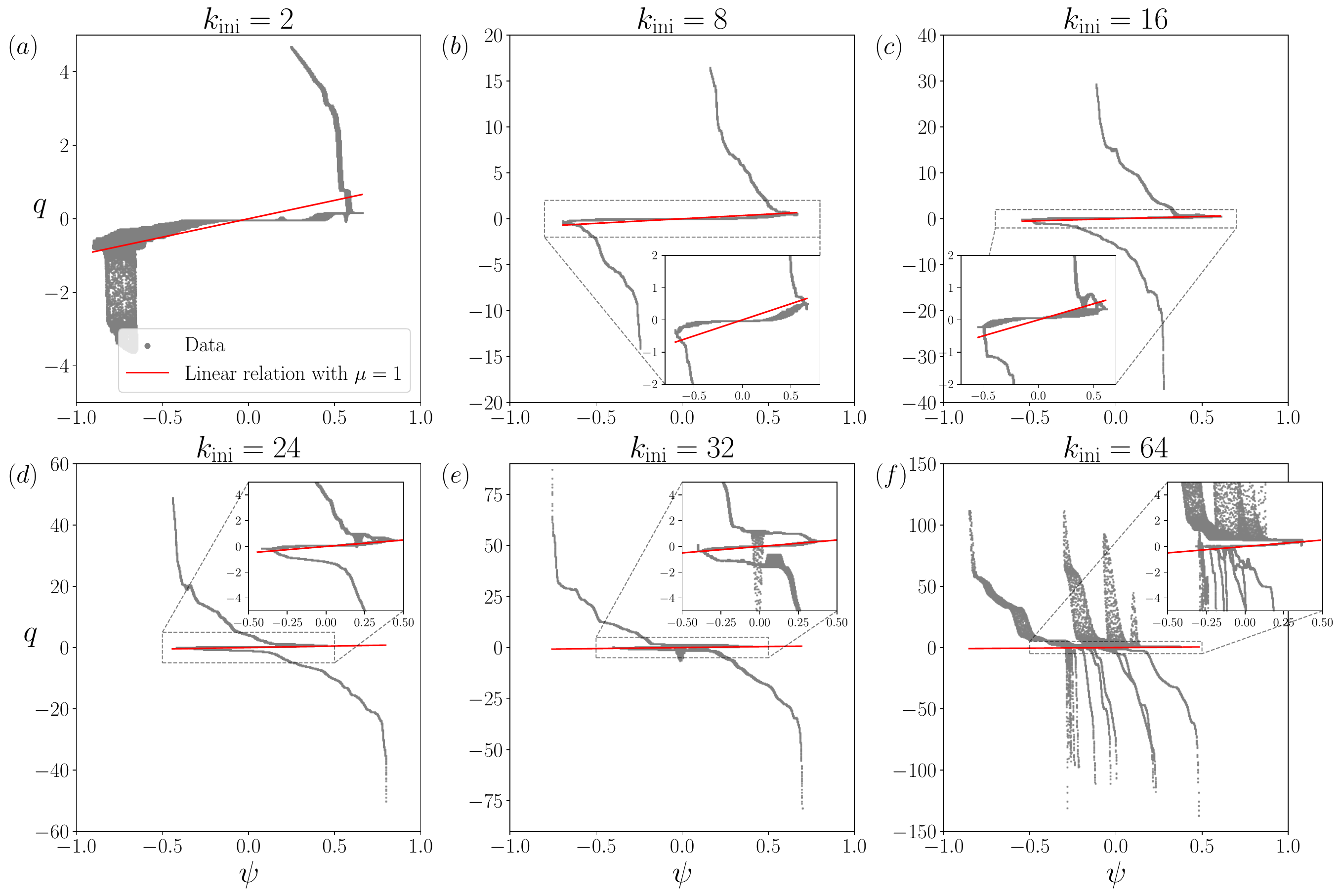}
    \caption{Scatter plots (in gray color) between $q$ and $\psi$ for different final states shown in Fig. \ref{fig:bump_contour}. 
    Red lines show the linear relation with the coefficient $\mu=1$ predicted by the minimum-enstrophy principle at the energy $E=0.25E_\#$.}
    \label{fig:bump_pv-psi}
\end{figure}

\begin{figure}
    \centering
    \includegraphics[width=1\textwidth]{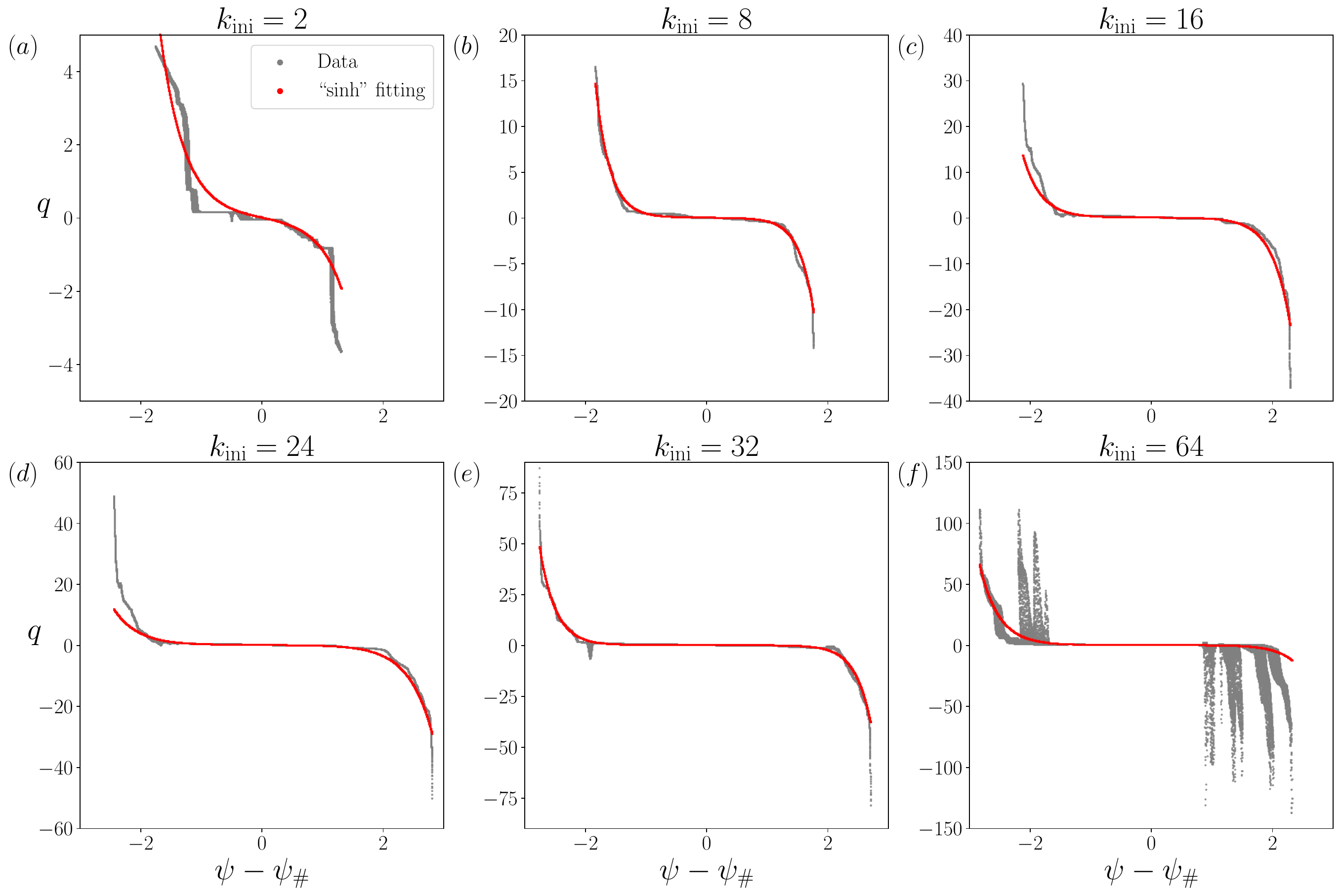}
    \caption{Scatter plots (in gray color) between $q$ and $\psi-\psi_\#$ for different final states shown in Fig. \ref{fig:bump_contour}.
    Red lines show the least squares fitting between $q$ and $\alpha\sinh{\beta(\psi-\psi_\#)}$.}
    \label{fig:bump_pv-psi-psic}
\end{figure}

We now examine the PV–streamfunction relationships in the final states. 
The corresponding scatter plots for the cases shown in Fig.~\ref{fig:bump_contour} are presented in Fig.~\ref{fig:bump_pv-psi}. 
Note that the scatter plots at different time instants nearly overlap, particularly for the quasi-stationary cases with $k_{\text{ini}} = 8$--$32$.
These scatter plots generally reveal three distinct branches, separated by two turning points that approximately correspond to the maximum and minimum values of the streamfunction.
The upper branch, located at the “top” of the scatter plots, features positive PV and a decreasing trend of streamfunction with increasing PV. 
According to geostrophic balance, this corresponds to increasingly lower pressure anomalies, and thus represents the intense cyclone at the center of the domain.
The lower branch, at the “bottom” of the plots, shows negative PV with a similar decreasing trend of streamfunction with increasing PV. 
This implies increasingly higher pressure anomalies, corresponding to the strong anticyclone located at the four corners of the domain.
The middle branch, which is sandwiched between the upper and lower branches, exhibits an increasing trend of the streamfunction with increasing PV, opposite to the behavior of the upper and lower branches.
This branch lies close to the linear PV–streamfunction relation predicted by the minimum-enstrophy state (indicated by red lines in Fig.~\ref{fig:bump_pv-psi}). 
The middle branch likely corresponds to regions away from the two dominant vortices, and is interpreted as representing the background flow.
As $k_{\text{ini}}$ increases, the vortices become more intense, as evidenced by the growing PV extrema and the widening range of streamfunction values between the turning points and the PV peaks. 
In the highest-wavenumber case ($k_{\text{ini}} = 64$), the scatter plot becomes more complex, exhibiting multiple branches—each likely corresponding to an individual cyclone or anticyclone observed in Fig.~\ref{fig:bump_contour}f.

Fig.~\ref{fig:bump_pv-psi} reveals a complex, non-monotonic relationship between PV and streamfunction in the final states. 
However, a more obvious functional relationship can be revealed through decomposition of the fields.
As shown in previous studies \cite{siegelmanTwodimensionalTurbulenceTopography2023,heMultipleStatesTwodimensional2024} and is evident in the middle branch of Fig.~\ref{fig:bump_pv-psi}, with the removal of vorticity outliers that are associated with the localized vortices, the background flow approximately satisfies a linear PV–streamfunction relation:
\begin{equation}
    q_b=\Delta\psi_b +\eta \sim \mu_b\psi_b,
\end{equation}
where $\mu_b$ is a linear factor characterizing the background flow.
Similarly, to investigate the functional relationship associated with the localized vortices (corresponding to the upper and lower branches in Fig.~\ref{fig:bump_pv-psi}), it is necessary to remove the background flow from the final state. This motivates us to examine the scatter plots between the residual PV, $q - q_b$, and the residual streamfunction, $\psi - \psi_b$.
However, since the value of $\mu_b$ varies across different final states and is unknown \textit{a priori}, we adopt a simplified approach by subtracting the critical background flow associated with a homogenized PV field, defined by $q_b = 0$ and $\psi_b = \psi_\#$ (i.e., $\mu_b = 0$). 
We will show that $\mu_b$ is small when the localized vortices are strong.
Then we remove only the topographic streamfunction $\psi_\#$, and inspect the scatter plots between $q$ and $\psi - \psi_\#$; the results are shown in Fig.~\ref{fig:bump_pv-psi-psic}.
Remarkably, these scatter plots exhibit a “$\sinh$”-like structure, suggesting that they may follow the empirical relation,
\begin{equation}\label{eq:empirical_relation}
    q \sim \alpha\sinh{\beta(\psi-\psi_\#)},
\end{equation}
where $\alpha$ and $\beta$ are fitting parameters. Following the classical analysis of 2D flat-bottom turbulence \cite{montgomeryRelaxationTwoDimensions1992}, we perform least-squares fitting to determine $\alpha$ and $\beta$ for each case. The fitted curves are shown in red in Fig.~\ref{fig:bump_pv-psi-psic}.
For the intermediate-wavenumber cases ($k_{\text{ini}} = 8$–$32$; Fig.~\ref{fig:bump_pv-psi-psic}$b$–$e$), where only two quasi-stationary vortices exist in the domain, the agreement between the data and the fitted curves is excellent, 
although discrepancies are visible along the upper, positive PV branches of the curves in Fig.~\ref{fig:bump_pv-psi-psic}$(c,d)$. The source of such discrepancies will be discussed later.
In the $k_{\text{ini}} = 2$ case (Fig.~\ref{fig:bump_pv-psi-psic}$a$), the “$\sinh$” structure is still discernible, but the fit is less accurate, likely due to the non-circularity of the vortices around the topographic extrema.
For the $k_{\text{ini}} = 64$ case (Fig.~\ref{fig:bump_pv-psi-psic}$f$), the presence of multiple vortex branches prevents a simple two-branch “$\sinh$” relation from capturing the full structure. Nevertheless, like-signed vortices still cluster toward the same end of the residual streamfunction $\psi - \psi_\#$, producing an overall ``$\sinh$''-like trend.

\begin{figure}
    \centering
    \includegraphics[width=0.8\linewidth]{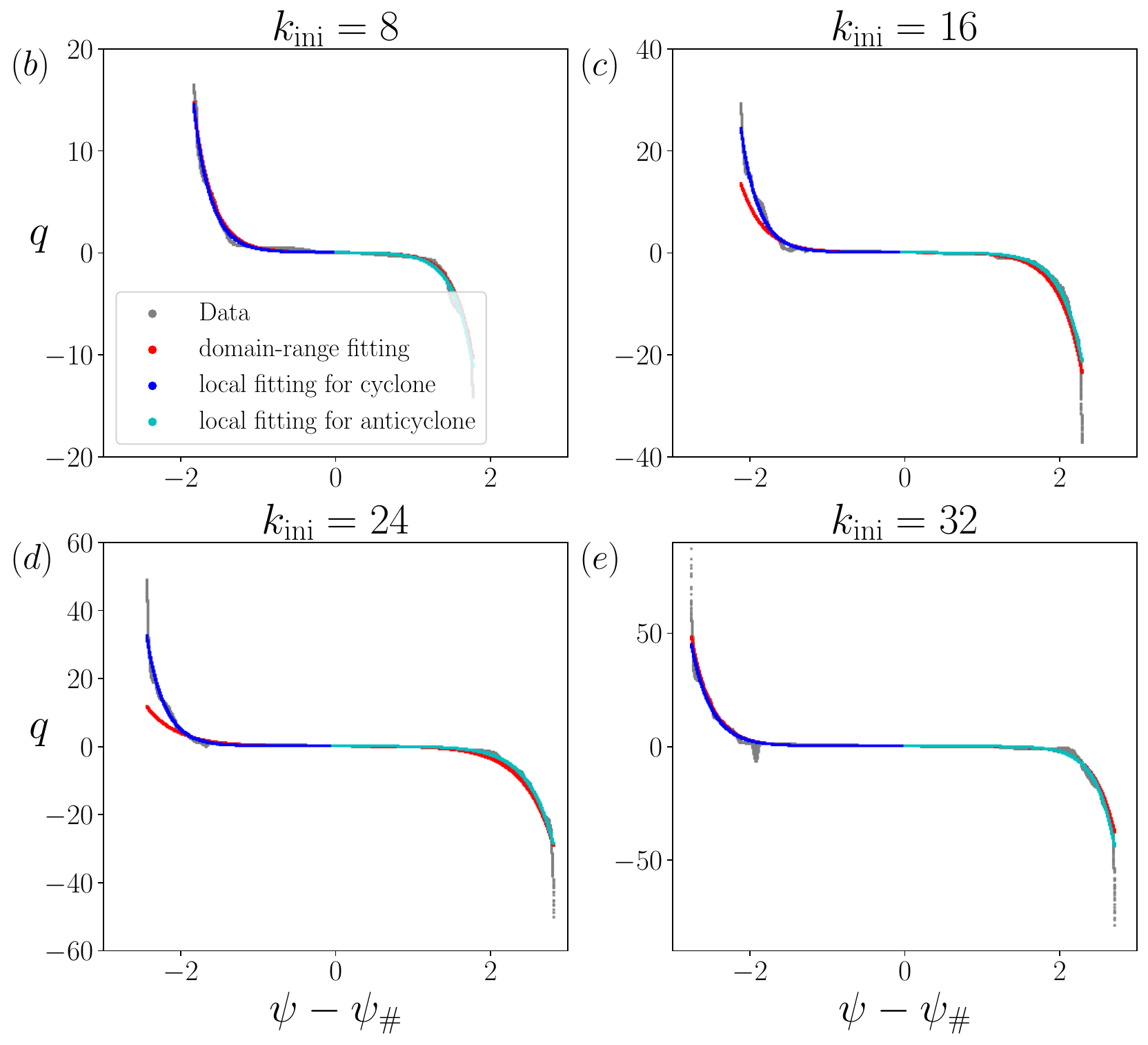}
    \caption{Comparisons of the domain-range fitting (red) with those performed for each vortex (blue: cyclone; cyan: anticyclone).}
    \label{fig:sinh_fitting_for_each_vortex}
\end{figure}

The above “$\sinh$” fittings are performed over the full domain such that all emergent vortices are encompassed, implicitly assuming that all vortices share the same fitting parameters.
To further assess this assumption, we conduct individual “$\sinh$” fittings for each vortex in the $k_{\text{ini}} = 8$–32 cases shown in Fig.~\ref{fig:bump_pv-psi-psic}$(b$–$e)$; the corresponding results are presented in Fig.~\ref{fig:sinh_fitting_for_each_vortex}.
The discrepancies along the positive PV branches resulting from the domain-scale fittings for the $k_{\text{ini}} = 16$ and $24$ cases (see Fig.~\ref{fig:bump_pv-psi-psic}$(c,d)$) are substantially reduced by the local, per‑vortex fittings.
This suggests that different vortices in certain cases—such as $k_{\text{ini}} = 16$ and $24$—may in fact correspond to distinct parameters in the “$\sinh$” functional relationship.
Moreover, the PV–streamfunction relationship can be multi‑valued, consistent with the findings reported in Ref.~\cite{solodochFormationAnticyclonesTopographic2021}.

In summary, after subtracting the background flow, the residual field associated with the localized vortices exhibits a $\sinh$-like behavior, reminiscent of those observed in flat-bottom turbulence.
In particular, the agreement is excellent for the intermediate-wavenumber cases ($k_{\text{ini}} = 8$–$32$; Fig.~\ref{fig:bump_pv-psi-psic}$b$–$e$), where vortices remain quasi-stationary. 
This indicates that stationary vortex solutions can be constructed for these cases by exploiting the ``$\sinh$'' relationship as follows.

\subsection{Empirical model of the stationary vortex solutions}
\label{sec:empirical_model}

Now we proceed to model the quasi-stationary final states as shown in  Fig.~\ref{fig:bump_contour}$(b–e)$.
Based on the above observations and previous studies \cite{siegelmanTwodimensionalTurbulenceTopography2023,heMultipleStatesTwodimensional2024}, we propose an empirical model as a superposition of a background flow and localized vortices:
\begin{equation}\label{eq:model}
    f^m = f^m_b + f^m_v,
\end{equation}
where $f \in [q, \psi, u, v]$. Here, $f^m_b$ and $f^m_v$ represent the modeled background flow and localized vortex components, respectively.

On one hand, the background flow is modeled using a linear PV–streamfunction relationship:
\refstepcounter{equation}\label{eq:model_background_flow}
$$
q_b^m = \mu_b \psi_b^m; \quad q_b^m = \Delta \psi_b^m + \eta. \eqno{(\theequation{\mathit{a},\mathit{b}})}
$$
For sinusoidal topography (Fig.\ref{fig:topography}), the background fields take explicit forms,
\begin{subequations}\label{eq:model_background_pv}
\begin{eqnarray}
    q_b^m(x, y, \mu_b) &=& \frac{\mu_b}{\mu_b + 1} (\cos{x} + \cos{y}), \\
    \psi_b^m(x, y, \mu_b) &=& \frac{1}{\mu_b + 1} (\cos{x} + \cos{y}), \\
    u_b^m(x,y,\mu_b) &=& \frac{1}{\mu_b+1}\sin{y}, \\
    v_b^m(x,y,\mu_b) &=& -\frac{1}{\mu_b+1}\sin{x},
\end{eqnarray}
\end{subequations}
for the PV, streamfunction, zonal and meridional velocities, respectively.

\begin{figure}
    \centering
    \includegraphics[width=1\linewidth]{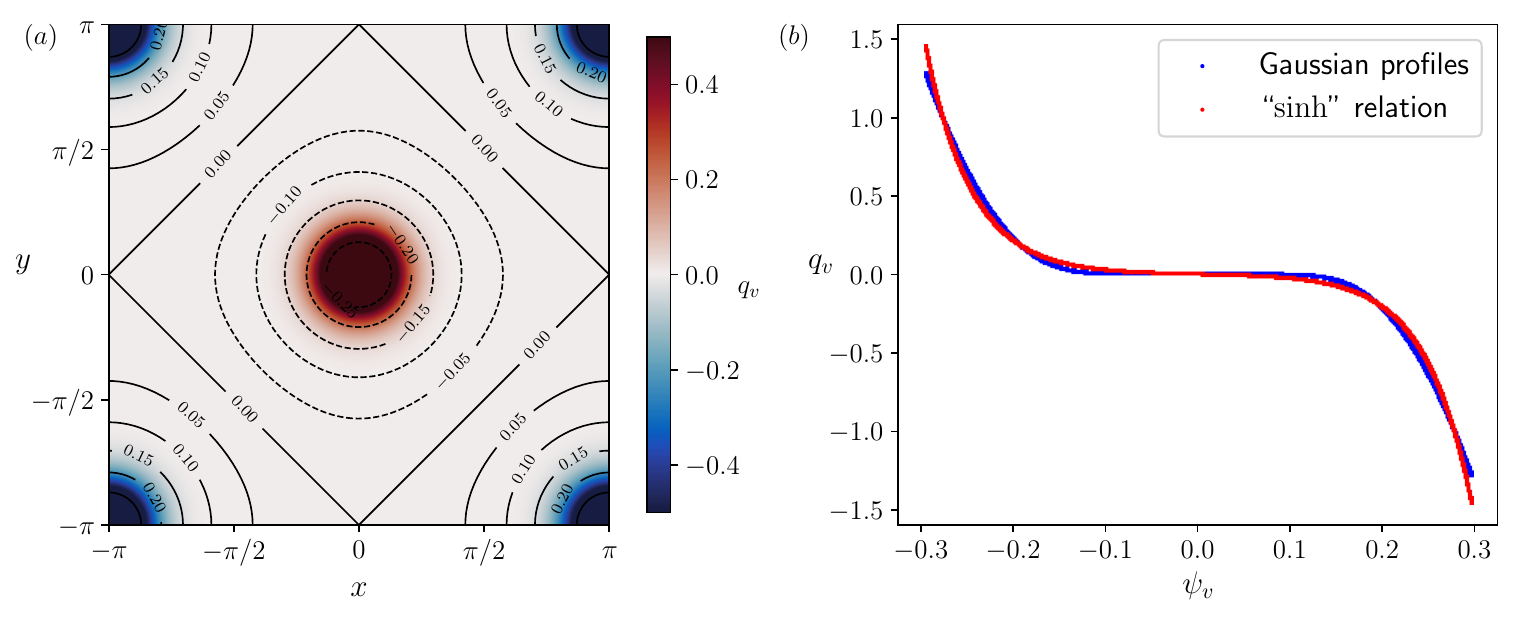}
    \caption{$(a)$ PV field ($q_v$, shown in pseudocolor) and streamfunction field ($\psi_v$, shown as contour lines), computed from equation (\ref{eq:model_vortices_pv}) with parameters $\Gamma_v = 1.0$ and $L_v = 0.5$;
    $(b)$ Scatter plot of $q_v$ versus $\psi_v$.}
    \label{fig:bump_qv-psiv}
\end{figure}

On the other hand, the localized vortices are assumed to follow a general nonlinear relationship:
\refstepcounter{equation}\label{eq:modeL_vortices}
$$
q_v^m = F(\psi_v^m); \quad q_v^m = \Delta \psi_v^m,
\eqno{(\theequation{\mathit{a},\mathit{b}})}
$$
where the function $F$ is generally nonlinear.
Based on our observations in \S~\ref{sec:q-psi_relation}, this function can be approximated by ``$\sinh$''.
However, solving the resulting nonlinear equation can be difficult.
To simplify the modeling, and in line with frequent observations in two-dimensional flat-bottom turbulence \cite{mcwilliamsVorticesTwodimensionalTurbulence1990,jimenezStructureVorticesFreely1996}, each localized vortex is instead represented using a Gaussian profile.
For the two dominant vortices observed in Fig.~\ref{fig:bump_contour}$(b$--$e)$, the PV field is modeled as a superposition of axisymmetric Gaussian profiles centered at topographic extrema, namely,
\begin{equation}\label{eq:model_vortices_pv}
    q_v^m(x,y,\Gamma_v,L_v) = q_{v1}^m(x,y,\Gamma_v,L_v) + q_{v2}^m(x,y,\Gamma_v,L_v)
\end{equation}
with
\begin{equation}\label{eq:model_vortices_cyclone}
    q_{v1}^m(x,y,\Gamma_v,L_v) = \frac{\Gamma_v}{\pi L_v^2}
    \mathrm{exp}\left[-\frac{x^2 + y^2}{L_v^2}\right]
\end{equation}
representing the cyclone centered at $(x, y) = (0, 0)$,
and
\begin{eqnarray}\label{eq:model_vortices_anticyclone}
    q_{v2}^m(x,y,\Gamma_v,L_v) &=& -\frac{\Gamma_v}{\pi L_v^2} \biggl\{
    \mathrm{exp}\left[-\frac{(x - \pi)^2 + (y - \pi)^2}{L_v^2}\right] +
    \mathrm{exp}\left[-\frac{(x - \pi)^2 + (y + \pi)^2}{L_v^2}\right] \nonumber\\
    &+& \mathrm{exp}\left[-\frac{(x + \pi)^2 + (y - \pi)^2}{L_v^2}\right] +
    \mathrm{exp}\left[-\frac{(x + \pi)^2 + (y + \pi)^2}{L_v^2}\right]
    \biggr\}
\end{eqnarray}
representing the anticyclone located at the four corners $(x, y) = (\pm\pi, \pm\pi)$.
For minimizing the number of fitting parameters, both vortices are assumed to have the same vortex strength $\Gamma_v(>0)$ and core length $L_v$.
The corresponding streamfunction and velocity fields do not have closed-form expressions, but can be obtained in spectral space.
It is worth noting that the vorticity and streamfunction fields of these two opposite-signed Gaussian profiles (\ref{eq:model_vortices_pv}) closely follow the ``$\sinh$'' relationship, as illustrated in Fig.~\ref{fig:bump_qv-psiv}.

The above formulas (\ref{eq:model}--\ref{eq:model_vortices_anticyclone}) constitute an empirical model for the quasi-stationary final states illustrated in Fig.~\ref{fig:bump_contour}($b$--$e$). 
Due to interactions between the background flow and localized vortices, this linear superposition does not yield an exact solution to the Jacobian $J(\psi,q)=0$.
However, near each topographic extremum, the background flow (\ref{eq:model_background_pv}) approximately behaves as
\begin{equation}
    q_b^m, \psi_b^m \sim r^2,
\end{equation}
where $r$ denotes the radial distance from the topographic extremum. In this regime, the contour lines of the background flow are circular and align with those of the axisymmetric Gaussian vortex located above the same topographic extremum.
As a result, the interaction between the background flow and the axisymmetric Gaussian vortex becomes negligible, and the empirical model can be regarded as an approximate local solution to the Jacobian $J(\psi,q)=0$ in the vicinity of each topographic extremum.

In the literature, vortices described by the vorticity formula (\ref{eq:model_vortices_cyclone}) are referred to as non-isolated vortices, characterized by nonzero circulations \cite{carnevaleEvolutionVortexStatistics1991,kloosterzielExperimentalStudyUnstable1991}. 
In contrast, another class of vortices, defined by the vorticity profile $(1-\tfrac{1}{2}r^2)e^{-\tfrac{1}{2}r^2}$ (where $r$ denotes the radial distance from the vortex center), are termed isolated vortices, which exhibit zero circulation \cite{hopfingerVorticesRotatingFluids1993,trielingDecayMonopolarVortices1998}. 
These isolated vortices are surrounded by an outer ring of oppositely signed vorticity. Their vorticity–streamfunction relationship is given by $\omega = 2\psi[1+\ln(2\psi)]$.
However, isolated vortices are unlikely to represent the localized vortices shown in Fig.~\ref{fig:bump_contour}, as no outer ring of oppositely signed vorticity is observed. 
Moreover, for large streamfunction values, the vorticity–streamfunction relationship of isolated vortices becomes approximately linear, which contradicts the super-linear behavior evident in Fig.~\ref{fig:bump_pv-psi-psic}.
Notably, Ref.~\cite{siegelmanPolarVortexCrystals2022} also employs non-isolated vorticity profiles to model vortex crystals in $\gamma$-plane turbulence. This strongly supports our use of non-isolated profiles, given that the $\gamma$ effect is dynamically equivalent to that from topography and that both studies adopt similar monoscale initial conditions to facilitate vortex generation.

The model (\ref{eq:model}), together with the background flow (\ref{eq:model_background_pv}) and vortex components (\ref{eq:model_vortices_pv}), contains three free parameters: the linear coefficient $\mu_b$ of the background flow, the vortex strength $\Gamma_v$, and the vortex core size $L_v$.
These parameters are determined by requiring that the integral quantities of the modeled state $f^m$ match those of the final state obtained from numerical simulations. Given three unknowns, we impose constraints on the energy, total enstrophy, and quartic Casimir of the modeled state $f^m$ by postulating:
\refstepcounter{equation}\label{eq:model_consraint}
$$
E^m(\mu_b,\Gamma_v, L_v) = E, \quad Q^m(\mu_b,\Gamma_v, L_v) = Q, \quad C_4^m(\mu_b,\Gamma_v, L_v) = C_4.
\eqno{(\theequation{\mathit{a},\mathit{b},\mathit{c}})}
$$
The left-hand sides of the constraints (\ref{eq:model_consraint}) are computed from the model $f^m$ and thus depend on the three parameters $\mu_b$, $\Gamma_v$, and $L_v$, while the right-hand sides are diagnosed from the final states of the numerical simulations.
By applying a nonlinear iterative solver to the system of constraints (\ref{eq:model_consraint}), the three parameters can be determined.

\begin{figure}
    \centering
    \includegraphics[width=0.7\textwidth]{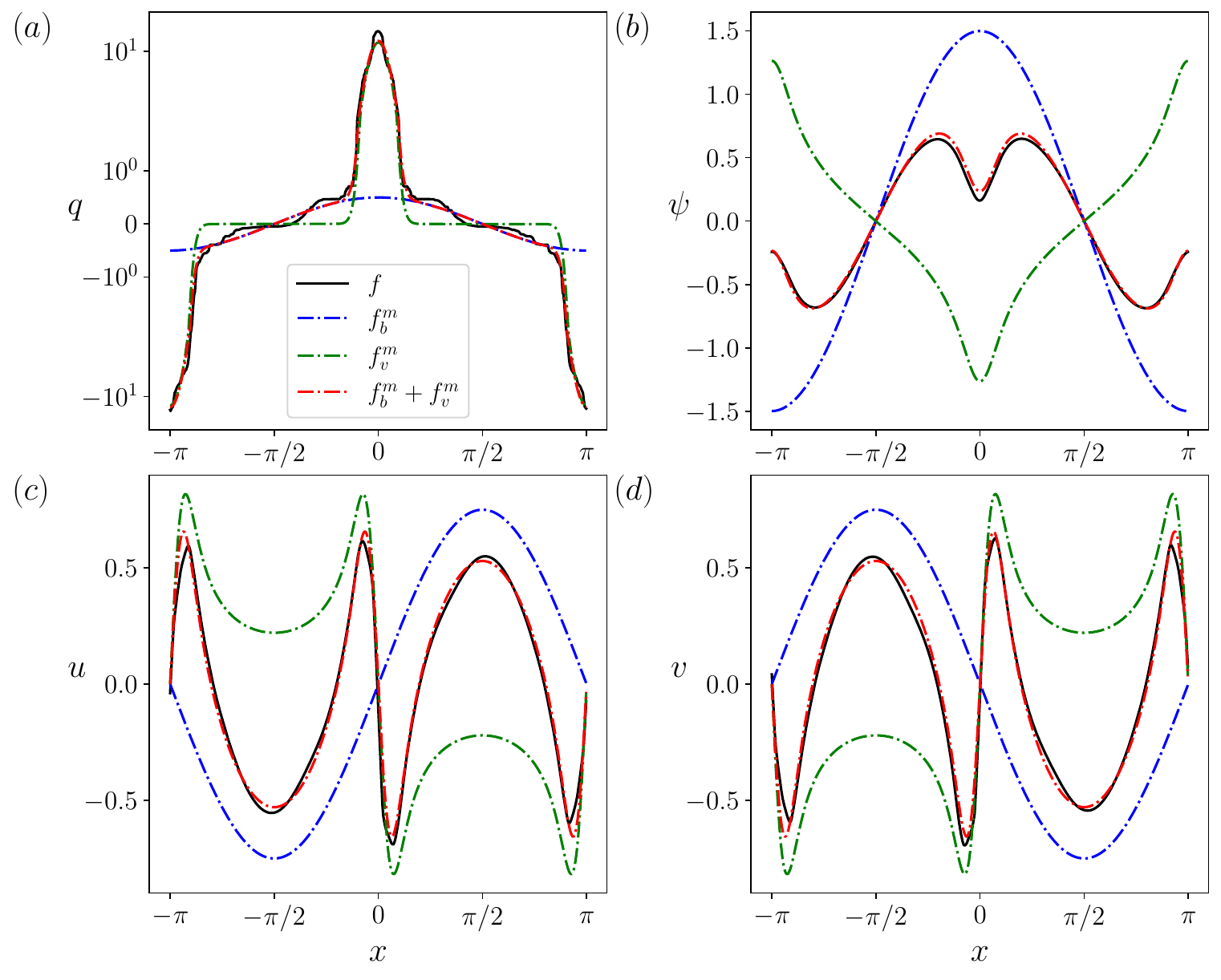}
    \caption{Comparisons of the PV ($a$), streamfunction ($b$), zonal velocity ($c$), and meridional velocity ($d$) along the dash dotted line shown in Fig. \ref{fig:bump_contour}$(a)$ between simulation data and empirical model for the final state corresponding to $k_{\text{ini}}=8$ (Fig. \ref{fig:bump_contour}$b$).
    $q$ in $(a)$ is shown in symlog scale in order to visualize the data within the range $[-1,1]$.}
    \label{fig:bump_modelling_kini=8}
\end{figure}

\begin{figure}
    \centering
    \includegraphics[width=0.7\textwidth]{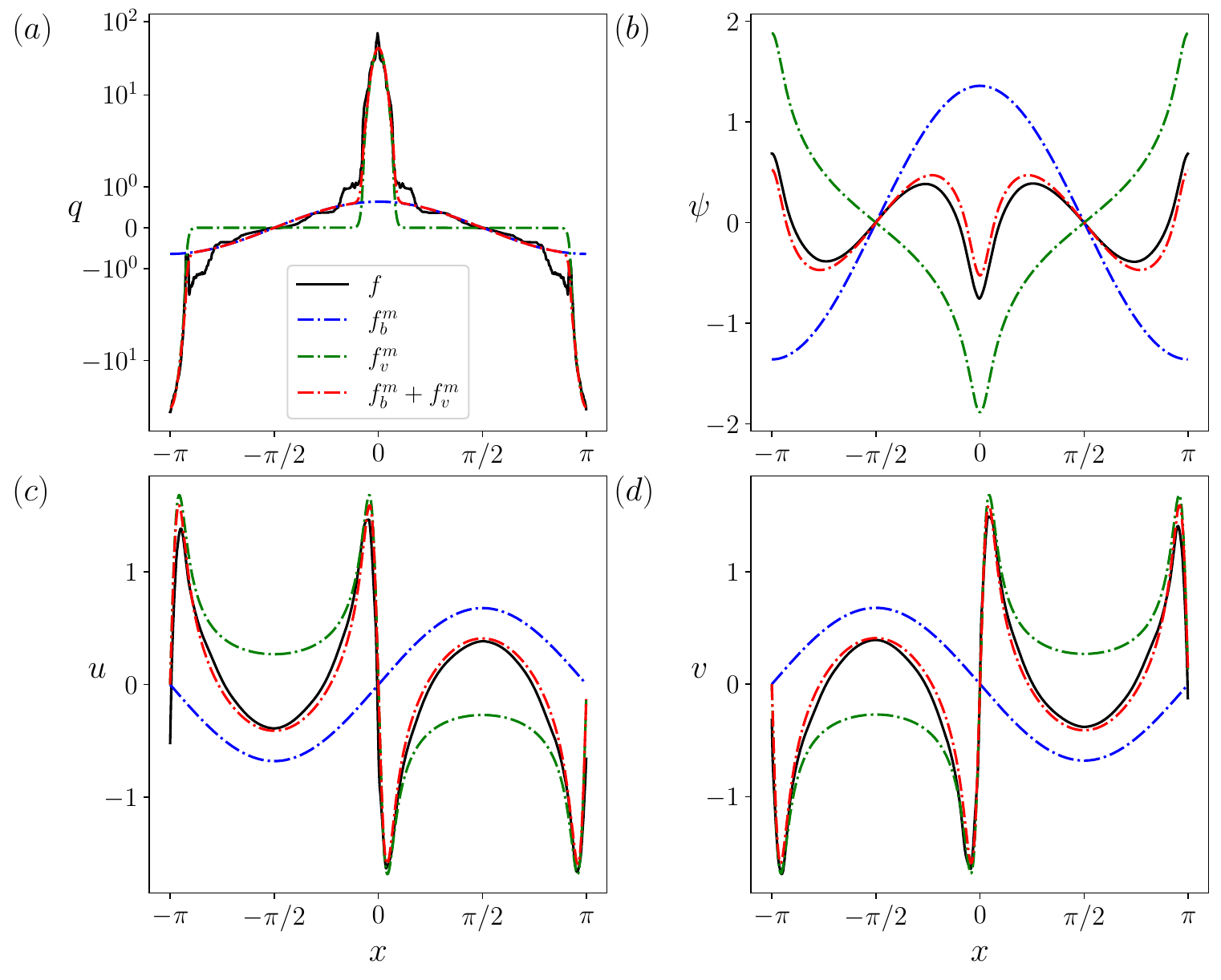}
    \caption{Same as Fig. \ref{fig:bump_modelling_kini=8} but for the final state corresponding to $k_{ini}=32$ (Fig. \ref{fig:bump_contour}$e$).}
    \label{fig:bump_modelling_kini=32}
\end{figure}

We now evaluate the performance of the superposition model (\ref{eq:model}), incorporating the background flow (\ref{eq:model_background_pv}) and localized vortices (\ref{eq:model_vortices_pv}), in reconstructing the final states.
Our assessment focuses on the intermediate $k_{\text{ini}}$ cases ($8$, $16$, $24$, $32$), shown in Fig.~\ref{fig:bump_contour}$(b$--$e)$, where the flow consists of a single cyclone and a single anticyclone. These vortices are axisymmetric and fully locked to their respective topographic extrema.
In Fig.~\ref{fig:bump_modelling_kini=8} and Fig.~\ref{fig:bump_modelling_kini=32}, we compare the final states from numerical simulations with those reconstructed using our model for the $k_{\text{ini}} = 8$ and $32$ cases, respectively. We present the profiles of potential vorticity $q$, streamfunction $\psi$, zonal velocity $u$, and meridional velocity $v$ along the diagonal of the domain (indicated by the dash-dotted line in Fig.~\ref{fig:bump_contour}$a$).
The results for $k_{\text{ini}} = 16$ and $24$ are similar and therefore omitted.

As shown in Fig.~\ref{fig:bump_modelling_kini=8}$(a)$ and Fig.~\ref{fig:bump_modelling_kini=32}$(a)$, the vortex model $q^m_v$ (green) successfully captures the peak and valley of the PV; the background model $q^m_b$ (blue) captures the general tendency in the transitional regions; and the superposition $q^m = q^m_b + q^m_v$ (red) effectively reproduces the overall spatial variation of PV.
However, the background model $q^m_b$ fails to capture smaller-scale PV structures, such as the PV staircases \cite{dritschel2008multiple} outside the peak and valley regions, as seen in Fig.~\ref{fig:bump_modelling_kini=8}$(a)$. 
As the vortices strengthen, these PV staircases become less pronounced, and the background model's performance improves outside the vortices (see Fig.~\ref{fig:bump_modelling_kini=32}$a$).
Despite its limitation in representing fine-scale PV structures, the superposition model adequately captures the large-scale structures of the streamfunction and velocity components, as demonstrated in Fig.~\ref{fig:bump_modelling_kini=8}$(b$--$d)$ and Fig.~\ref{fig:bump_modelling_kini=32}$(b$--$d)$.

It is important to note that the flow direction of the background field is opposite to that of the vortices, as illustrated by the streamfunction and velocity profiles in Fig.~\ref{fig:bump_modelling_kini=8}$(b$--$d)$ and Fig.~\ref{fig:bump_modelling_kini=32}$(b$--$d)$, consistent with the distinct correlations of the background flow and vortices with the underlying topography.

\begin{figure}
    \centering
    \includegraphics[width=1\linewidth]{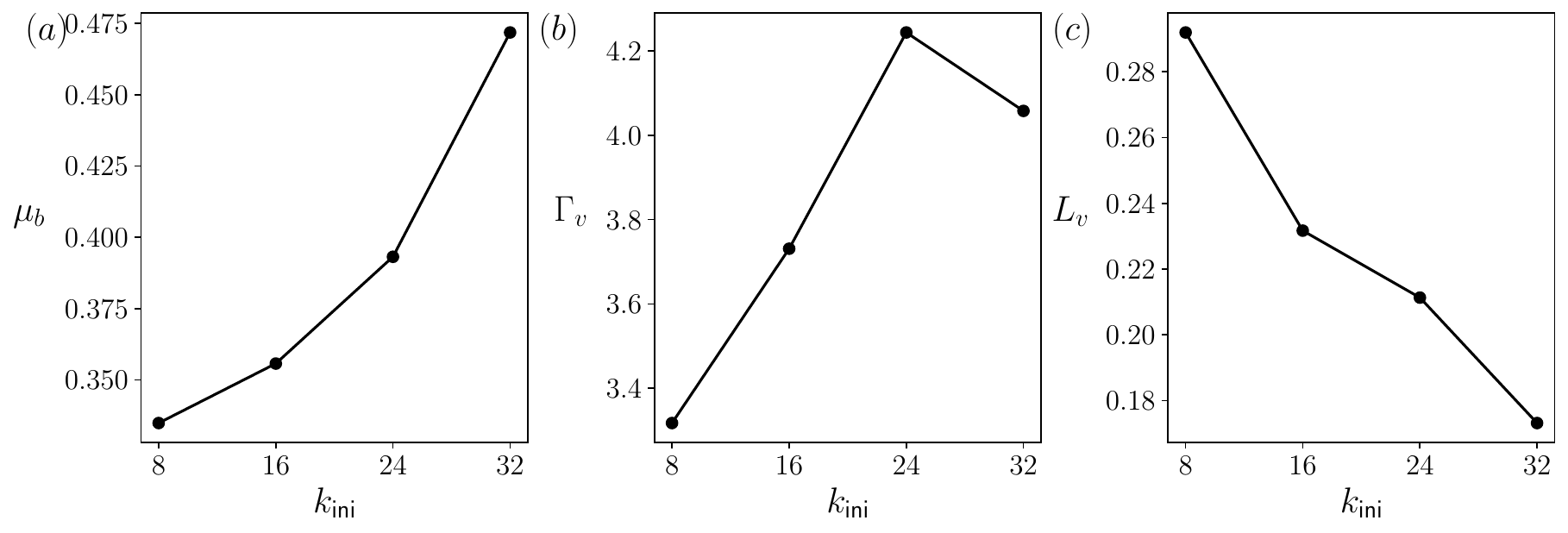}
    \caption{Model parameters $\mu_b$, $\Gamma_v$, and $L_v$ for different $k_{\text{ini}}$.}
    \label{fig:bump_model_parameters}
\end{figure}

In Fig.~\ref{fig:bump_model_parameters}, we present the model parameters for different $k_{\text{ini}}$ cases.
For the background model, the linear coefficient $\mu_b$ is consistently less than $1$, which corresponds to the value of the minimum-enstrophy solution at the energy level of our simulations. 
This observation aligns with previous findings on background flows after the removal of vorticity outliers \citep{siegelmanTwodimensionalTurbulenceTopography2023,heMultipleStatesTwodimensional2024}; the presence of vortices tends to homogenize the background PV, thereby reducing the linear coefficient $\mu_b$ of the background flow.
However, the observed increase in $\mu_b$ with $k_{\text{ini}}$ in the range $[8, 32]$ contradicts earlier results obtained over random topography \citep{heMultipleStatesTwodimensional2024}, where higher $k_{\text{ini}}$ was expected to seed more vortices, enhance PV mixing, and thus lower $\mu_b$. Despite this discrepancy, the $k_{\text{ini}} = 64$ case—featuring multiple like-signed vortices—exhibits a more homogenized background PV compared to the $k_{\text{ini}} = 8$--$32$ cases (see Fig.~\ref{fig:bump_contour}), consistent with previous findings over random topography \citep{heMultipleStatesTwodimensional2024}.
We speculate that in the $k_{\text{ini}} = 8$--$32$ cases, the presence of only two vortices within the domain may be insufficient to drive significant PV homogenization.
On the other hand, as $k_{\text{ini}}$ increases, the vortex strength $\Gamma_v$ generally increases (Fig.~\ref{fig:bump_model_parameters}$b$), while the vortex core size $L_v$ decreases (Fig.~\ref{fig:bump_model_parameters}$c$), indicating that the vortices become stronger.

\begin{figure}
    \centering
    \includegraphics[width=1\linewidth]{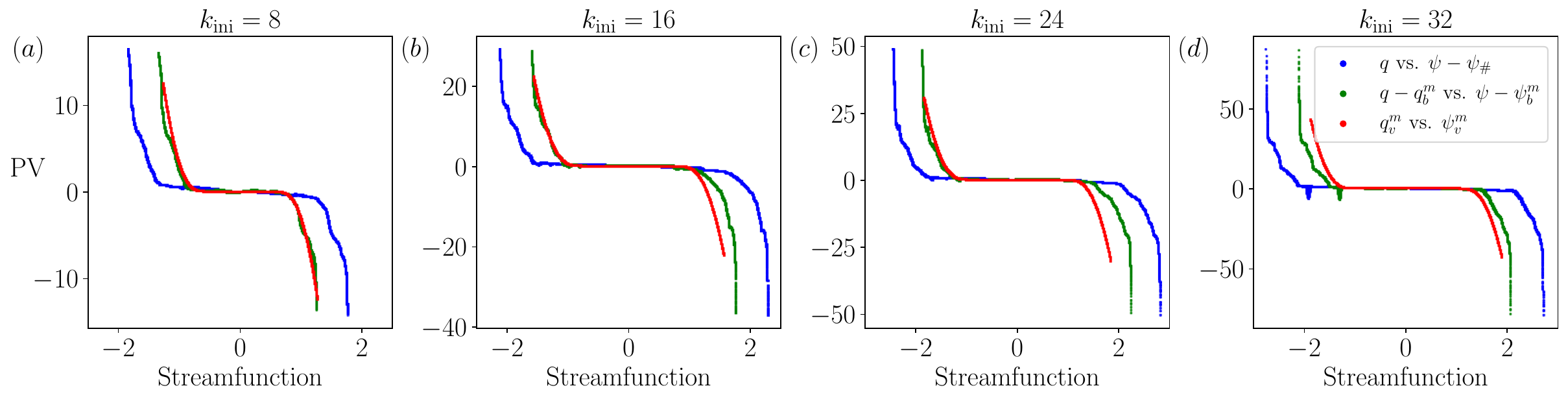}
    \caption{Scatter plots of $q$ vs. $\psi-\psi_\#$ (removal of topographic streamfunction corresponding to homogeneous PV; blue color), $q-q_b^m$ vs. $\psi-\psi_b^m$ (removal of the background model; green color) and $q_v^m$ vs. $\psi_v^m$ (Gaussian vortex model; red color).}
    \label{fig:q-psi_three-relations}
\end{figure}

When examining the scatter plots for localized vortices in \S~\ref{sec:q-psi_relation}, we removed only the topographic streamfunction $\psi_\#$ corresponding to the homogeneous PV ($\mu_b = 0$) from the simulation data for all cases, since the background linear factor $\mu_b$ was unknown at that stage. 
Within the modeling framework, however, $\mu_b$ is determined to be non-zero and varies across cases, as shown in Fig.~\ref{fig:bump_model_parameters}($a$). 
At this point, the background model for $q_b^m$ and $\psi_b^m$ can be constructed and then removed from the simulation data.
In Fig.~\ref{fig:q-psi_three-relations}, the scatter plots of $q$ versus $\psi - \psi_\#$ with the homogeneous PV field removed (blue) are compared to those of $q - q_b^m$ versus $\psi - \psi_b^m$ with the background model removed (green). 
Both sets exhibit the same trend, following the ``$\sinh$'' relationship, which confirms the validity of removing the homogeneous PV field in \S~\ref{sec:q-psi_relation}. 
This similarity arises because the background PV is small relative to the PV of the vortices and can effectively be treated as zero (i.e., homogeneous PV). Although the topographic streamfunction $\psi_\#$ associated with homogeneous PV is larger in magnitude than the actual background streamfunction, the resulting over-shifts in the scatter plots do not alter the functional trend.

Figure~\ref{fig:q-psi_three-relations} also presents the scatter plots of $q_v^m$ versus $\psi_v^m$ for the Gaussian vortex model (red). 
These plots generally align with, or lie close to, those of $q - q_b^m$ versus $\psi - \psi_b^m$ (green), confirming the validity of representing localized vortices as Gaussian vortices. 
Some discrepancies, however, are evident. 
On the one hand, the Gaussian vortex model fails to capture the PV extrema; nevertheless, this limitation does not significantly affect its ability to approximate the streamfunction and velocity fields (Figs.~\ref{fig:bump_modelling_kini=8}–\ref{fig:bump_modelling_kini=32}). 
On the other hand, the model performs better on one branch than the other, as the cyclone actually differs from the anticyclone, whereas our Gaussian vortex model presumes equal strength and core length for both to reduce the number of adjustable parameters.

\subsection{Applicability to random topography and high energy levels}
\label{sec:random_topography_and_high_energy}

Thus far, our analysis has focused on 2D turbulence above an idealized large-scale sinusoidal topography and at low energy levels, which allow simple spatial distributions of localized vortices in the quasi-stationary states. 
A natural question arises as to how the proposed framework performs in the presence of more complex topography or at higher energy levels. Complex topography can host a vast number of vortices with varying polarities, sizes and strengths, distributed at local topographic extrema in an unpredictable way. 
At elevated energy levels, vortices are more vigorous and mobile. 
In these circumstances, we do not aim to model the global flow fields, as the intricate spatial distributions of vortices introduce numerous parameters requiring calibration, and a mere superposition of two stationary solutions cannot capture the unsteady behavior of vortices. 
Nevertheless, it remains feasible to examine two fundamental assumptions underlying our vortex modeling approach: the validity of the “$\sinh$” relationship and the Gaussian-shaped vortex profiles. 
In this subsection, we will investigate whether localized vortices under these more challenging conditions adhere to the two assumptions.

\subsubsection{Random topography}

\begin{figure}
    \centering
    \includegraphics[width=0.9\linewidth]{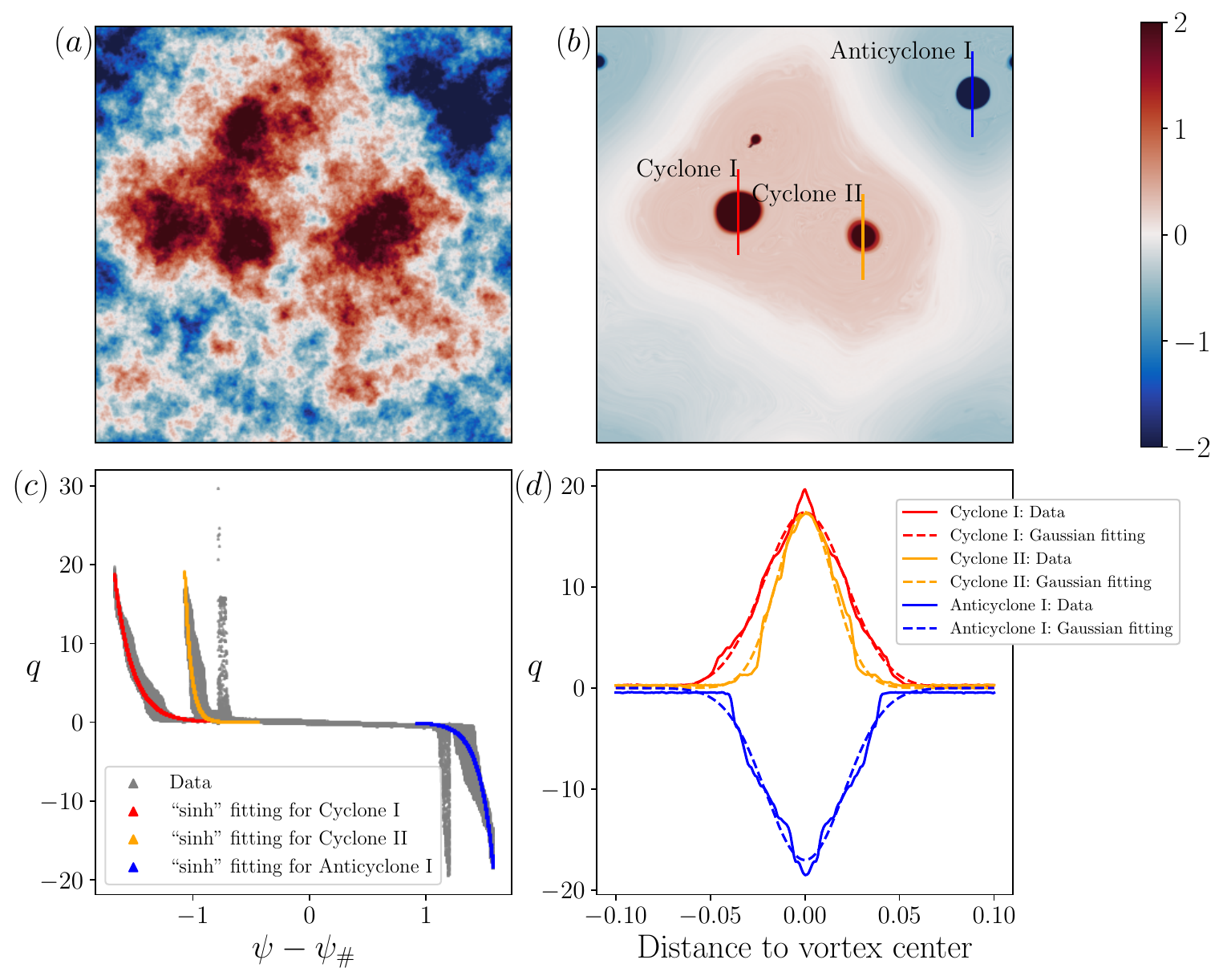}
    \caption{Results for a run above a random topography and initialized with the energy level $E=0.25E_\#$ and the wavenumber $k_{\text{ini}}=16$: 
    ($a$) Topographic PV $\eta$;
    ($b$) PV $q$ after a long time;
    ($c$) Scatter plots between PV $q$ and residual streamfunction $\psi-\psi_\#$ and local ``$\sinh$'' fitting for three dominant vortices shown in ($b$);
    ($d$) PV profiles of the vortices on the cuts shown in ($b$).}
    \label{fig:random_topography}
\end{figure}

We begin by examining low-energy solutions over a random multi-scale topography, utilizing numerical data from our previous work \cite{heMultipleStatesTwodimensional2024}. 
The multi-scale topography is expressed as
\begin{equation}
    \eta(x,~y) = \sum_{\bk}\eta_{\bk}\rme^{\rmi\bk\cdot\boldsymbol{x}}.
\end{equation}
The Fourier amplitude takes the form $\eta_{\bk}=\alpha\rme^{\rmi\phi_{\bk}}k^{-3/2}$, with $\alpha$ as the normalization factor and $\phi_{\bk}$ as the random phase.
This construction yields a multi-scale topography characterized by a $k^{-2}$ power spectrum.
Additional details regarding the simulations conducted over this multi-scale topography can be found in Ref. \cite{heMultipleStatesTwodimensional2024}.
The results for one typical state above the multi-scale topography are presented in Fig.~\ref{fig:random_topography}. 
Panel ($a$) shows the topographic PV $\eta(x,y)$ corresponding to a random topography. 
Panel ($b$) displays a PV snapshot $q$ after a long-time integration, initialized with a low energy level $E = 0.25E_\#$ and wavenumber $k_{\mathrm{ini}} = 16$. 
Due to the complexity of the random topography, multiple cyclones and anticyclones of varying sizes and strengths emerge over arbitrary local topographic extrema. 
Panel ($c$) presents the scatter plot of PV $q$ against the residual streamfunction $\psi - \psi_\#$ and the local ``$\sinh$'' fittings for three dominant vortices identified in panel ($b$). 
The three dominant vortices follow the ``$\sinh$'' trend.
Finally, panel ($d$) shows the PV profiles of the three dominant vortices. Least-squares fitting based on Gaussian profiles indicates that the vortex structures are well-approximated by Gaussian shapes.

\begin{figure}
    \centering
    \includegraphics[width=1\linewidth]{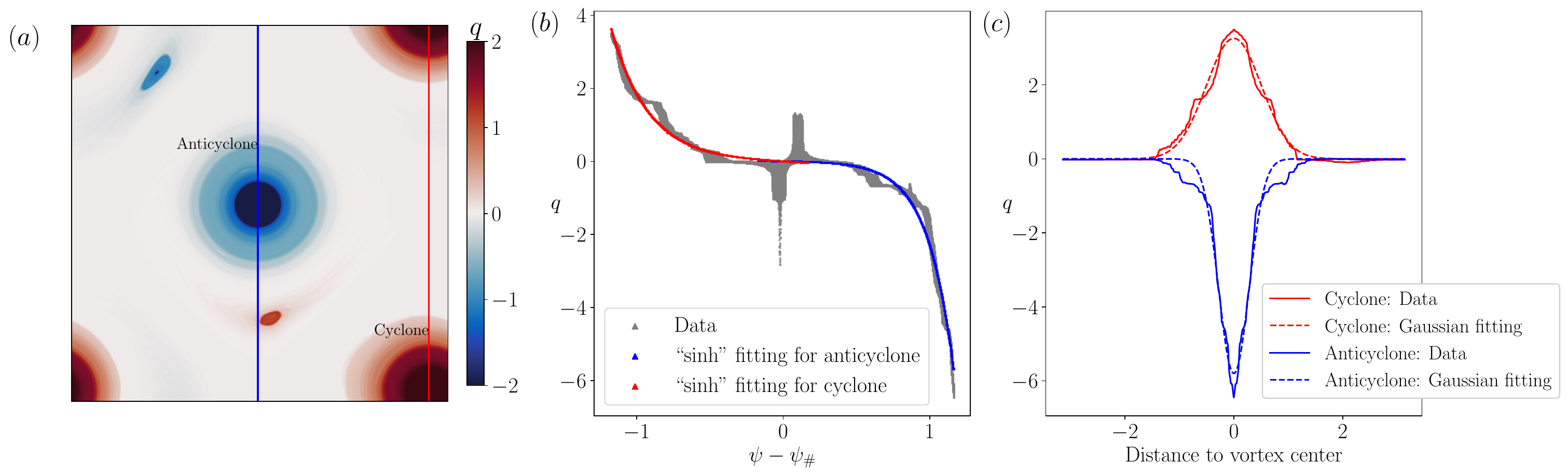}
    \caption{Results for a run above the topography (see Fig.\ref{fig:topography}) initialized with the energy level $E=2E_\#$ and the wavenumber $k_{\text{ini}}=1$:
    ($a$) PV $q$ at the time instant 5000;
    ($b$) Scatter plots of PV $q$ against residual streamfunction $\psi-\psi_\#$ and local ``$\sinh$'' fitting for the two vortices shown in ($a$);
    ($c$) PV profiles of the vortices on the sections shown by blue and read lines in ($a$).}
    \label{fig:high_energy_large_scale}
\end{figure}

\begin{figure}
    \centering
    \includegraphics[width=1\linewidth]{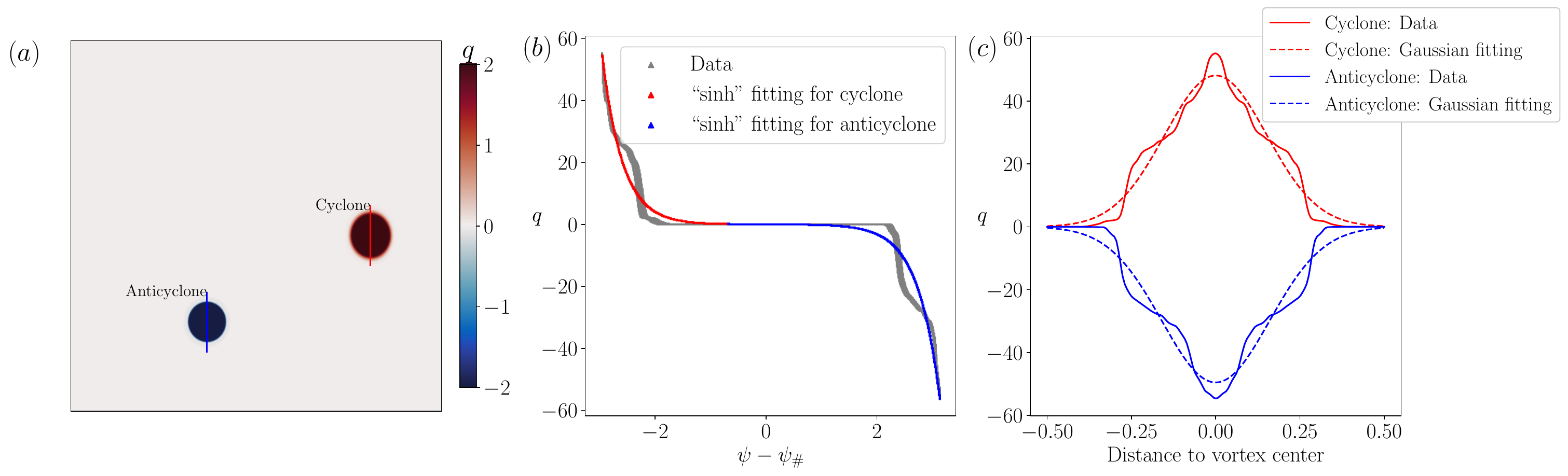}
    \caption{
    Similar to Fig.\ref{fig:high_energy_large_scale} but for a run initialized with the wavenumber $k_{\text{ini}}=12$.
    }
    \label{fig:high_energy_small_scale}
\end{figure}

\subsubsection{High energy levels}

We next turn to the high energy levels.
We perform numerical simulations above the idealized large-scale sinusoidal topography (Fig.\ref{fig:topography}) initialized at the energy level $E=2E_\#$ and two different initial wavenumbers $k_{ini}=1$ and $k_{ini}=12$; the results are shown in Figs.\ref{fig:high_energy_large_scale} and \ref{fig:high_energy_small_scale}, respectively.
Similar to those above a random topography in our previous work \cite{heMultipleStatesTwodimensional2024}, the long-term states at high energy levels display two different regimes.
On the one hand, as shown in Fig.\ref{fig:high_energy_large_scale}$(a)$, when the length scale of the initial field is large and comparable with the domain scale, the anticyclone and cyclone are locked to the topographic maximum and minimum, respectively, exactly opposite to the correlation at the low energy level (see Fig.\ref{fig:bump_contour}).
However, the locking is ``slack'', as the vortices display strong unsteadiness. 
They oscillate around their respective topographic extrema, with their motion confined to local regions surrounding these extrema.
The background flow in this case was found to possess a negative linear factor \cite{heMultipleStatesTwodimensional2024}.
On the other hand, as shown in Fig.\ref{fig:high_energy_small_scale}$(a)$, when the initial length scale is small, the movements of vortices are much less confined, and they are roaming within a homogenized background PV field.
Additional high-energy simulation results in the Appendix~\ref{sec:appendix} further demonstrate the dependence on the initial wavenumber and the robustness of the vortex-topography anti-correlation across different high energy levels.

As before, we perform least square fittings to inspect whether the ``$\sinh$'' relation and the Gaussian shapes are satisfied by the vortices in these two regimes.
The results are shown in Figs.~\ref{fig:high_energy_large_scale}$(b$--$c)$ and \ref{fig:high_energy_small_scale}$(b$--$c)$, respectively.
Both the “$\sinh$” relationship and the Gaussian vortex profiles remain clearly identifiable even at high energy levels, demonstrating the robustness of these two features for vortices across different energy regimes of two‑dimensional topographic turbulence.

\section{Barotropic stability}
\label{sec:stability}

When constructing the empirical model for the stationary vortex solutions that can be compared with numerical simulations, we leveraged the correlation between vortices and topography observed in numerical simulations at low energy levels. Specifically, we positioned a cyclone and an anticyclone above the topographic elevation and depression, respectively (see Eqs. \ref{eq:model_vortices_pv}–\ref{eq:model_vortices_anticyclone}). 
In this section, we aim to interpret this vortex–topography correlation at low energy levels through linear stability analyses of the stationary vortex solutions. 
As a related outcome, the vortex–topography anti-correlation observed at high energy levels (see Fig. \ref{fig:high_energy_large_scale}$(a)$) can also be interpreted within the same theoretical framework.

\subsection{Equations}

As discussed earlier, the empirical model for stationary vortex solutions (\ref{eq:model}, \ref{eq:model_background_pv}, \ref{eq:model_vortices_pv}, \ref{eq:model_vortices_cyclone}, \ref{eq:model_vortices_anticyclone})—a linear superposition of a background flow and Gaussian vortices—is not a global exact solution to the Jacobian $J(\psi,q)=0$, and therefore does not permit global linear stability analyses. 
However, the streamlines of the two components align near the topographic extrema, making their sum a local exact solution to $J(\psi,q)=0$. 
This allows us to perform linear stability analyses of the local approximate solutions around each topographic extremum. 
Here, we focus on the elevation at the domain center; the reverse conclusions apply to the depression at the four corners of the topography.

Around the topographic maximum (Fig.~\ref{fig:topography}), upon Taylor expansion, the approximate form of the topographic PV is circular and given by
\begin{equation}\label{eq:circular_elevation}
\eta(r) \approx 2 - \frac{r^2}{2}
\end{equation}
in polar coordinates $(r,\theta)$ centered at the topographic maximum.
The quadratic variation of this topographic PV from the center again indicates its equivalence to the $\gamma$-effect imposed by a polar cap studied in Ref.~\cite{siegelmanPolarVortexCrystals2022}. The base flow above this circular elevation (\ref{eq:circular_elevation}) is obtained by taking the approximate form of the empirical model for stationary vortex solutions (\ref{eq:model}, \ref{eq:model_background_pv}, \ref{eq:model_vortices_pv}, \ref{eq:model_vortices_cyclone}, \ref{eq:model_vortices_anticyclone}) at the topographic maximum. The PV and azimuthal velocity of the base flow are thus given by
\begin{equation}\label{eq:base_pv}
Q(r) =  \frac{\mu_b}{\mu_b+1} \left(2-\frac{r^2}{2}\right) + \frac{\Gamma_v}{\pi L_v^2}\exp\left(-\frac{r^2}{L_v^2}\right)
\end{equation}
and
\begin{equation}\label{eq:base_velocity}
U_\theta(r) = -\frac{1}{\mu_b+1}r + \frac{\Gamma_v}{2\pi r}\left[1-\exp\left(-\frac{r^2}{L_v^2}\right)\right],
\end{equation}
respectively.
The first terms in Eqs.~(\ref{eq:base_pv}–\ref{eq:base_velocity}) correspond to the topographic flow, derived from the leading terms of the Taylor expansion of Eq.~(\ref{eq:model_background_pv}) at the domain center, while the second terms represent a Gaussian vortex. 
Following previous studies \cite{benilovStabilityTwoLayerQuasigeostrophic2005,gonzalezLinearStabilityMonopolar2023}, we perform linear stability analyses of the axisymmetric base flow (\ref{eq:base_pv}–\ref{eq:base_velocity}) on a circular disk. 
The stability depends on the background linear factor $\mu_b$, vortex strength $\Gamma_v$, and vortex core length $L_v$. 
The signs of $\mu_b$ and $\Gamma_v$ are unrestricted; we examine the stability of vortices with different polarities (depending on $\Gamma_v$) in the presence of topographic flows of varying energy levels (depending on $\mu_b$). 
Several limiting cases are noteworthy:
\begin{itemize}
    \item When $\mu_b \to +\infty$, the background flow reduces to the circular topography $\eta(r)$ in Eq. (\ref{eq:circular_elevation}), and only the stability of the Gaussian vortex over topography is considered.
    \item When $\mu_b = 0$, the background PV becomes homogenized, and we study the stability of a Gaussian vortex in a uniform PV environment.
    \item When $\Gamma_v = 0$, the analysis focuses solely on the stability of the background topographic flow.
\end{itemize}

To perform linear stability analysis, we decompose the flow field into the base flow and a perturbation,
\begin{equation}\label{eq:flow_decomp}
    [q,\psi,u_r,u_\theta](r,\theta,t) = [Q(r),\Psi(r),0,U_\theta(r)] + [q',\psi',u_r',u_\theta'](r,\theta,t).
\end{equation}
Plugging the decomposition (\ref{eq:flow_decomp}) into the governing equation (\ref{eq:qg_equation}) and omitting the viscous and nonlinear terms, we obtain the linearized equation for the perturbation,
\begin{equation}
    \left(\frac{\partial}{\partial t}+\frac{U_\theta}{r}\frac{\partial}{\partial\theta}\right)\Delta\psi-\frac{1}{r}\frac{d Q}{d r}\frac{\partial\psi}{\partial\theta} = 0,
\end{equation}
where the prime $'$ denoting the perturbation is omitted without ambiguity.
The perturbation is assumed to take a normal form,
\begin{equation}
    \psi=\hat{\psi}(r)\mathrm{exp}\left[\mathrm{i}(m\theta-\omega t)\right],
\end{equation}
where $m$ is the azimuthal wavenumber, $\Re{\omega}$ is the frequency and $\Im{\omega}$ is the growth rate.
We then obtain an eigenvalue problem for $\omega$,
\begin{equation}\label{eq:eigenvalue_problem}
    \left(-\omega+m\frac{U_\theta}{r}\right)\left(\frac{\partial^2}{\partial r^2}+\frac{1}{r}\frac{\partial}{\partial r}-\frac{m^2}{r^2}\right)\hat{\psi}-m\frac{1}{r}\frac{d Q}{d r}\hat{\psi} =0.
\end{equation}
The boundary conditions for the eigenfunction $\hat{\psi}(r)$ are taken to be null at the origin and infinity \cite{gentInstabilityBarotropicCircular1986}:
\begin{equation}\label{eq:eigevalue_boundary_condition}
    \hat{\psi}=0 \quad \text{at} \quad r=0; \quad \hat{\psi}\rightarrow 0 \quad \text{for} \quad r\rightarrow+\infty.
\end{equation}

\subsection{Conditions for barotropic instability}

\begin{figure}
    \centering
    \includegraphics[width=0.5\linewidth]{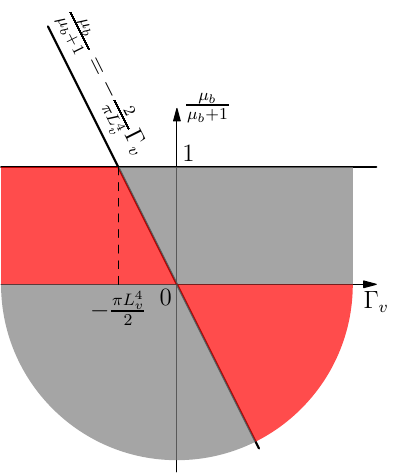}
    \caption{Unstable (red) and stable (gray) regions in the ${\mu_b}/{(\mu_b+1)}\text{---}\Gamma_v$ plane.}
    \label{fig:unstable_region}
\end{figure}

According to Rayleigh’s theorem for QG flows above topography \cite{gonzalezLinearStabilityMonopolar2023}, the necessary condition for barotropic instability is that the PV gradient of the base flow vanishes at some radial distance $r_z$:
\begin{equation}\label{eq:condition_for_barotropic_instability}
    \frac{dQ}{dr}\Big|_{r=r_z}=0.
\end{equation}
Differentiating the base flow PV in (\ref{eq:base_pv}) with respect to $r$ yields
\begin{equation}\label{eq:dQ/dr}
    \frac{dQ}{dr} = r\left[-\frac{\mu_b}{\mu_b+1} - \frac{2\Gamma_v}{\pi L_v^4}\mathrm{exp}\left(-\frac{r^2}{L_v^2}\right)\right].
\end{equation}
Then, the condition (\ref{eq:condition_for_barotropic_instability}) is satisfied if  the base flow parameters satisfy the inequalities
\begin{equation}
\begin{cases}
-\dfrac{2}{\pi L_v^4}\Gamma_v < \dfrac{\mu_b}{\mu_b+1} < 0, & \Gamma_v > 0; \\[6pt]
-\dfrac{2}{\pi L_v^4}\Gamma_v > \dfrac{\mu_b}{\mu_b+1} > 0, & \Gamma_v < 0; \\[6pt]
\dfrac{\mu_b}{\mu_b+1} \leq 1.
\end{cases}
\end{equation}
The last inequality for ${\mu_b}/{(\mu_b+1)}$ results from $\mu_b>-1$. 
We treat ${\mu_b}/{(\mu_b+1)}$ and $\Gamma_v$ as two governing parameters of linear stability.
The straight line $\mu_b/(\mu_b+1)=-2/(\pi L_v^4)\Gamma_v$ (with $-2/(\pi L_v^4)$ as the slope) and the horizontal axis $\mu_b/(\mu_b+1)=0$ delineate the region of $\mu_b/(\mu_b+1)\leq1$.
The unstable and stable regions in the ${\mu_b}/{(\mu_b+1)}\text{---}\Gamma_v$ plane are shown in red and gray colors in Fig. \ref{fig:unstable_region}, respectively.

The stable regions in the first and third quadrants in the ${\mu_b}/{(\mu_b+1)}\text{---}\Gamma_v$ plane (Fig. \ref{fig:unstable_region}) are directly relevant to the relationships between vortices and topography observed in simulations. 
These regions satisfy $\mu_b \cdot \Gamma_v > 0$, indicating that the background linear factor $\mu_b$ and the vortex strength $\Gamma_v$ have the same sign.
Because the topography considered here is an elevation (\ref{eq:circular_elevation}), a cyclone ($\Gamma_v > 0$) and an anticyclone ($\Gamma_v < 0$) remain stable above this elevation at low ($\mu_b > 0$) and high ($\mu_b < 0$) background energy levels, respectively.
For a topographic depression, the stability pattern reverses owing to the sign reversal of the topographic PV (\ref{eq:circular_elevation}) and thereby the first term of the PV gradient (\ref{eq:dQ/dr}). 
These stability arguments are therefore consistent with, and help explain, the vortex–topography correlation and anti‑correlation observed in the low‑ and high‑energy simulations (see Figs.~\ref{fig:bump_contour} and \ref{fig:high_energy_large_scale}).

The unstable regions in the ${\mu_b}/{(\mu_b+1)}\text{---}\Gamma_v$ plane (Fig.~\ref{fig:unstable_region}) appear as subsets of the second and fourth quadrants, where the condition $\mu_b \cdot \Gamma_v < 0$ holds. However, this condition is only necessary—not sufficient—for instability, because additional stable regions exist within these quadrants, specifically in the area bounded by the line $\mu_b/(\mu_b+1) = -2/(\pi L_v^4)\Gamma_v$ and the vertical axis ($\Gamma_v = 0$).
As will be shown later, for a vortex length $L_v$ typical of those observed in simulations, the dividing line $\mu_b/(\mu_b+1) = -2/(\pi L_v^4)\Gamma_v$ nearly coincides with the vertical axis. Consequently, the stable pockets in the second and fourth quadrants become negligible. In this regime, it is reasonable to characterize the unstable regions simply by the condition $\mu_b \cdot \Gamma_v < 0$.
Thus, anticyclones ($\Gamma_v < 0$) and cyclones ($\Gamma_v > 0$) can become unstable above the topographic elevation at low ($\mu_b > 0$) and high ($\mu_b < 0$) background energy levels, respectively.

\subsection{Numerical results of the eigenvalue problem}

\begin{figure}
    \centering
    \includegraphics[width=1\linewidth]{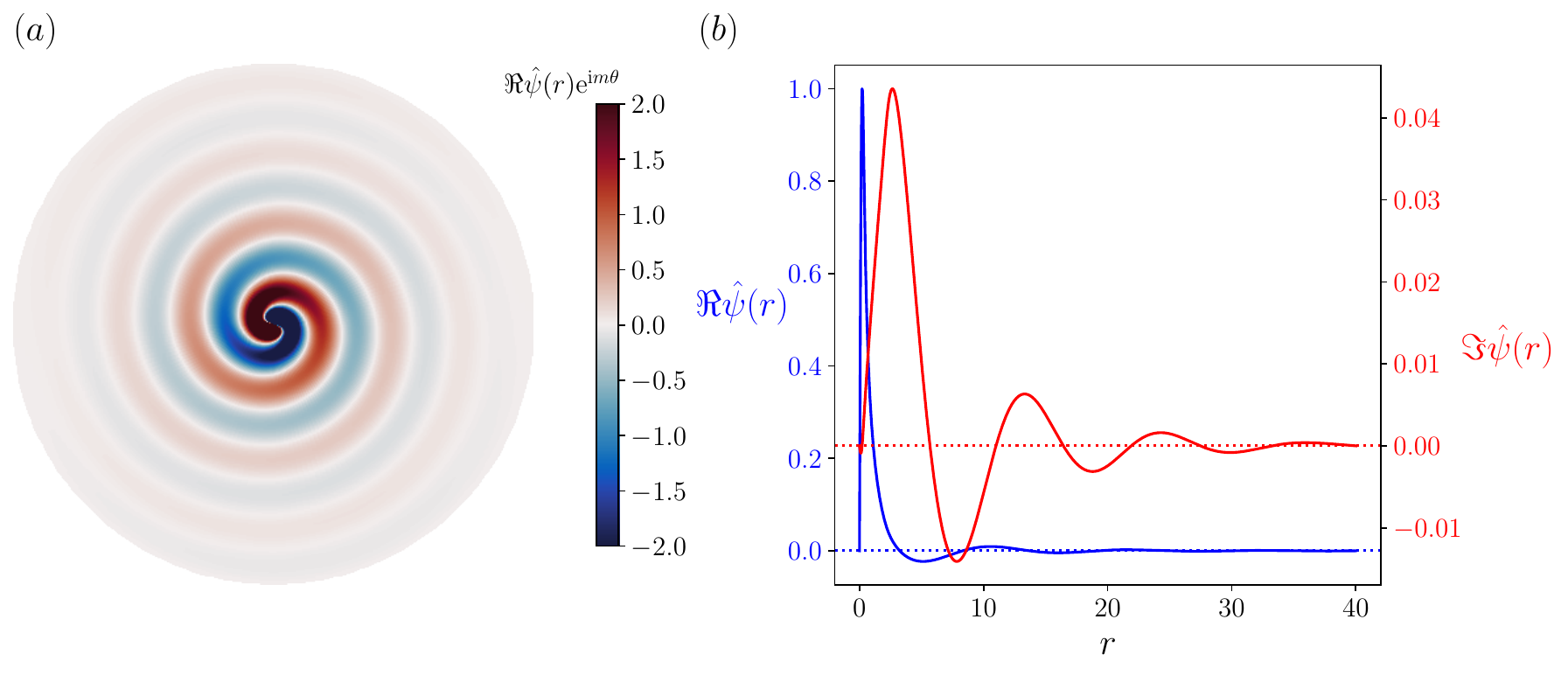}
    \caption{Shape of the eigenfunction $\hat{\psi}(r)$ for an unstable mode with azimuthal wavenumber $m=1$ and eigenvalue $\omega \approx -2.7906 + 0.9315\mathrm{i}$, obtained for the base-flow parameters $(\mu_b = +\infty, \Gamma_v = -100, L_v = 0.2)$.
($a$) Pseudocolor plot of the real part of $\hat{\psi}(r)\mathrm{e}^{\mathrm{i} m \theta}$ on the disk showing the spiral decay.
($b$) Radial distributions of the real and imaginary parts of $\hat{\psi}(r)$ normalized such that the entry with the largest amplitude is set as $1$.}
    \label{fig:eigenfunction}
\end{figure}

To validate the stability arguments derived from Rayleigh’s theorem, we numerically solve the eigenvalue problem (\ref{eq:eigenvalue_problem}–\ref{eq:eigevalue_boundary_condition}) using the open‑source software \texttt{Dedalus}, which employs spectral methods \cite{burnsdedalus2020}. Since all coefficients in the formulation are real‑valued, the eigenmodes are either unstable or neutrally stable. For each set of base‑flow parameters, we compute the full spectrum for a given azimuthal wavenumber $m$ and identify the most unstable mode, defined as the one with the largest growth rate $\Im{\omega}$.
Although the boundary condition (\ref{eq:eigevalue_boundary_condition}) is formally imposed at infinity, the computations are performed on a finite disk of radius $R$, with a homogeneous boundary condition $\hat{\psi}(R)=0$ applied at $r=R$. The radial direction is discretized using $N$ grid points. Through trial and error, we select $R=40$ and $N=2000$ for all computations. This choice ensures that, for unstable modes, the eigenfunction $\hat{\psi}(r)$ decays fully when reaching the outer boundary (see Fig.~\ref{fig:eigenfunction}); for neutrally stable modes, the numerical error of the growth rates measured by the ratio $|\Im{\omega}|/|\Re{\omega}|$ remains below $10^{-6}$. 
Further increasing either $R$ or $N$ does not qualitatively alter the results.

A series of eigenvalue problems are solved by varying the base‑flow parameters $(\mu_b, \Gamma_v, L_v)$. The vortex length is fixed at $L_v = 0.2$, a representative value for the vortices observed in our previous simulations (see Fig.~\ref{fig:bump_model_parameters}$c$). The background linear factor $\mu_b$ is specified by prescribing values of $\mu_b/(\mu_b + 1)$ in the interval $[1, -1]$, corresponding to $\mu_b \in [+ \infty, -0.5]$. The vortex strength $\Gamma_v$ is varied through the range $[-100, 100]$ with a incremental strength of $1$. Note that the values of $(\mu_b, \Gamma_v)$ in our previous empirical modeling for turbulence states (see Fig.~\ref{fig:bump_model_parameters}$a,b$) are fully covered by the above chosen parameter ranges.
For the small vortex length $L_v = 0.2$, the dividing line
$\mu_b/(\mu_b + 1) = -2/(\pi L_v^4)\Gamma_v$,
which separates the stable and unstable regions in the $\mu_b/(\mu_b+1)\text{---}\Gamma_v$ plane (Fig.~\ref{fig:unstable_region}), nearly coincides with the vertical axis. 
Thus, the combinations of chosen $\mu_b/(\mu_b+1)$ and $\Gamma_v$ fall cleanly into the unstable region when $\mu_b \cdot \Gamma_v < 0$ and into the stable region when $\mu_b \cdot \Gamma_v > 0$.

\begin{figure}
    \centering
    \includegraphics[width=0.7\linewidth]{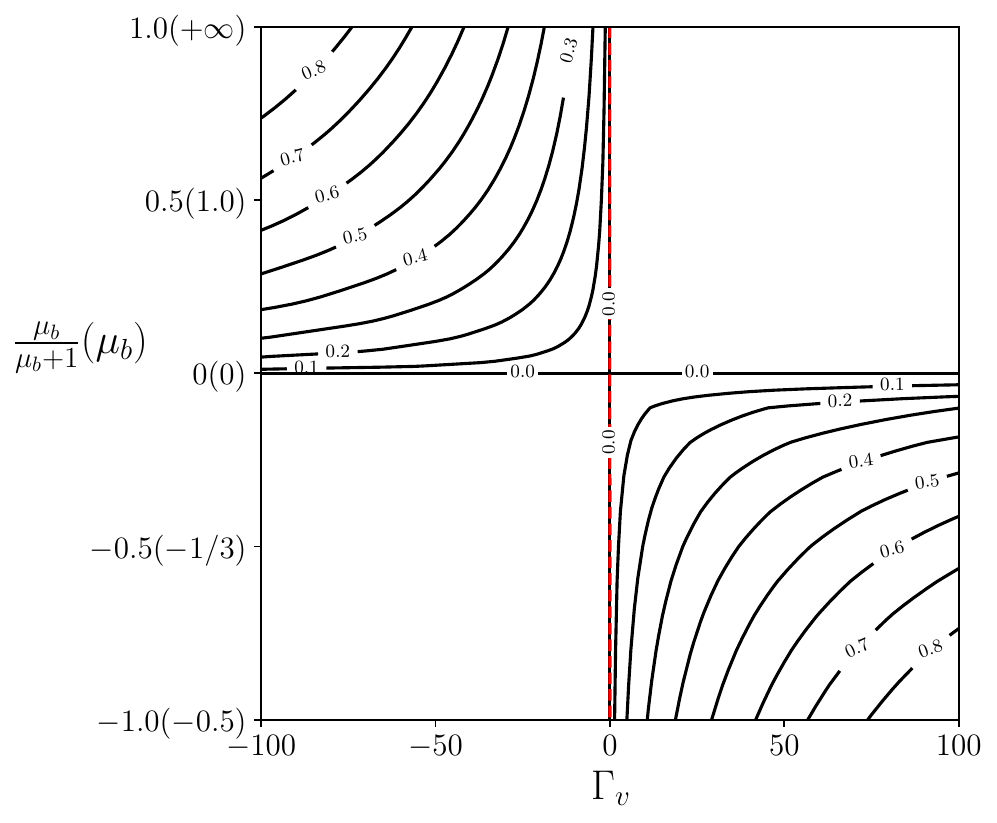}
    \caption{Contour plot of the growth rate $\Im{\omega}$ in the $\mu_b/(\mu_b+1)\text{---}\Gamma_v$ plane for fixed parameters $L_v = 0.2$ and $m = 1$.
    The red dashed line is $\mu_b/(\mu_b+1)=-2/(\pi L_v^4)\Gamma_v$.
    The brackets of the tick labels in the vertical axis show the values of $\mu_b$.}
    \label{fig:growth_rate_contour}
\end{figure}

Across all parameter combinations considered, the most unstable mode is consistently found at the azimuthal wavenumber $m = 1$, which is therefore analyzed in detail as follows. 
The contour map of the growth rate $\Im{\omega}$ in the $\mu_b/(\mu_b+1)\text{---}\Gamma_v$ plane is shown in Fig.~\ref{fig:growth_rate_contour}. 
Instability occurs when $\mu_b \cdot \Gamma_v < 0$, whereas stability occurs when $\mu_b \cdot \Gamma_v > 0$, consistent with Rayleigh’s criterion. 
For the special case of $\mu_b = 0$, corresponding to a homogenized background PV, a Gaussian vortex is always stable regardless of its polarity. When the Gaussian vortex is absent ($\Gamma_v = 0$), the topographic flow is also stable, in agreement with earlier studies \cite{carnevaleNonlinearStabilityStatistical1987}.
In summary, our numerical results confirm that, for a topographic elevation, cyclones ($\Gamma_v > 0$) and anticyclones ($\Gamma_v < 0$) are stable at low ($\mu_b > 0$) and high ($\mu_b < 0$) background energy levels, respectively, while vortices of opposite polarity are unstable under these conditions. 
For a topographic depression, the stability characteristics reverse. 
These patterns are once again consistent with, and help explain, the vortex–topography correlations and anti‑correlations observed in numerical simulations.

\section{Conclusion}
\label{sec:conclusion}

The final states of freely decaying two-dimensional (2D) topographic turbulence are characterized by the coexistence of a background topographic flow and localized vortices. 
The background flow is found to satisfy a linear potential vorticity (PV)–streamfunction relation \cite{siegelmanTwodimensionalTurbulenceTopography2023,heMultipleStatesTwodimensional2024}, whereas the detailed structure of the vortices remains less well understood. 
In this study, we characterize these localized vortices by drawing on insights from flat-bottom turbulence. 
For oceanic relevance and analytical tractability, we focus on quasi-stationary final states over a large-scale sinusoidal topography with one bump and one dip, where a single cyclone and anticyclone  emerge above the respective topographic extrema.

Our analysis reveals that, once the topographic streamfunction associated with the background flow is removed, the residual fields of these vortices exhibit a “$\sinh$”-like PV–streamfunction relation, particularly in quasi-stationary states. 
This functional form is commonly observed for vortices in flat-bottom turbulence, suggesting a universal behavior despite the restricted mobility imposed by topography.
Furthermore, motivated by the Gaussian profiles commonly associated with vortices in flat-bottom cases, we propose an empirical model for the quasi-stationary states of 2D topographic turbulence. 
The model combines a background flow satisfying a linear PV–streamfunction relation with Gaussian vortices centered at the topographic extrema. 
Near these extrema, the streamlines of the vortices and the background flow are nearly parallel, yielding an approximately stationary solution to the inviscid QG PV equation. 
Model parameters—including the background linear factor, vortex strength, and core length—are determined by matching integral invariants of energy, total enstrophy and quartic Casimir from simulations. 
This model successfully reproduces the domain-scale variations of PV, streamfunction and velocity fields. 
Although our quantitative analysis focuses on an idealized topography and a low energy level, we confirm that the key assumptions—the “$\sinh$”-like trend and Gaussian vortex profiles—remain overall applicable to complex random topography and high-energy regimes. 
These findings provide explicit stationary vortex solutions consistent with the final states of 2D topographic turbulence and incorporate local vortex structures for the first time.

The final states of freely decaying 2D topographic turbulence exhibit characteristic correlations between vortices and topography. 
At low energy levels, cyclones and anticyclones are locked to the respective topographic elevation and depression (Fig.~\ref{fig:bump_contour}). 
At high energy levels, when the initial conditions are dominated by domain-scale structures, the opposite configuration arises: anticyclones align with elevations and cyclones with depressions (Fig.~\ref{fig:high_energy_large_scale}). 
These contrasting vortex–topography correlations across energy regimes can be explained through linear stability analyses of our stationary vortex solutions. 
When the background flow is in the low-energy regime, the cyclone/elevation (or anticyclone/depression) configuration is stable, whereas the anticyclone/elevation (or cyclone/depression) configuration is unstable. 
In the high-energy regime of the background flow, the stability properties reverse, a process that has not been reported before. 
These vortex stability and instability characteristics in the presence of a background topographic flow are consistent with the correlations observed in numerical simulations across different energy levels.

In this work, we focused on topographic turbulence within a barotropic QG framework. Extending this approach to baroclinic QG systems is a natural next step, as the real ocean is stratified. 
Indeed, regimes of topographic turbulence have been reported in numerical simulations of two-layer baroclinic turbulence over topography \cite{pudig2025baroclinic}. 
However, the detailed structure of localized vortices in such baroclinic settings remains unexplored. 
According to Ref.~\cite{benilovStabilityTwoLayerQuasigeostrophic2005}, in a two-layer QG system, a bottom elevation tends to stabilize cyclones and destabilize anticyclones, whereas the opposite holds for a depression. 
This suggests that the vortex–topography correlations identified in this study may persist in baroclinic turbulence above topography, albeit with additional complexity introduced by stratification. 
Investigating  the potential existence of localized vortices and the pertinent vortex-topography correlations in baroclinic systems will be an important direction for future work.

\begin{acknowledgments}
This work is supported by the Research Grants Council (RGC) of Hong Kong under award General Research Fund 16307324, and by the Center for Ocean Research (CORE), a joint research center between Laoshan Laboratory and HKUST.
The authors are grateful to two anonymous referees whose constructive comments significantly improved this work.
\end{acknowledgments}

\appendix

\section{Additional high-energy simulation results}
\label{sec:appendix}

\begin{figure}
    \centering
    \includegraphics[width=1\linewidth]{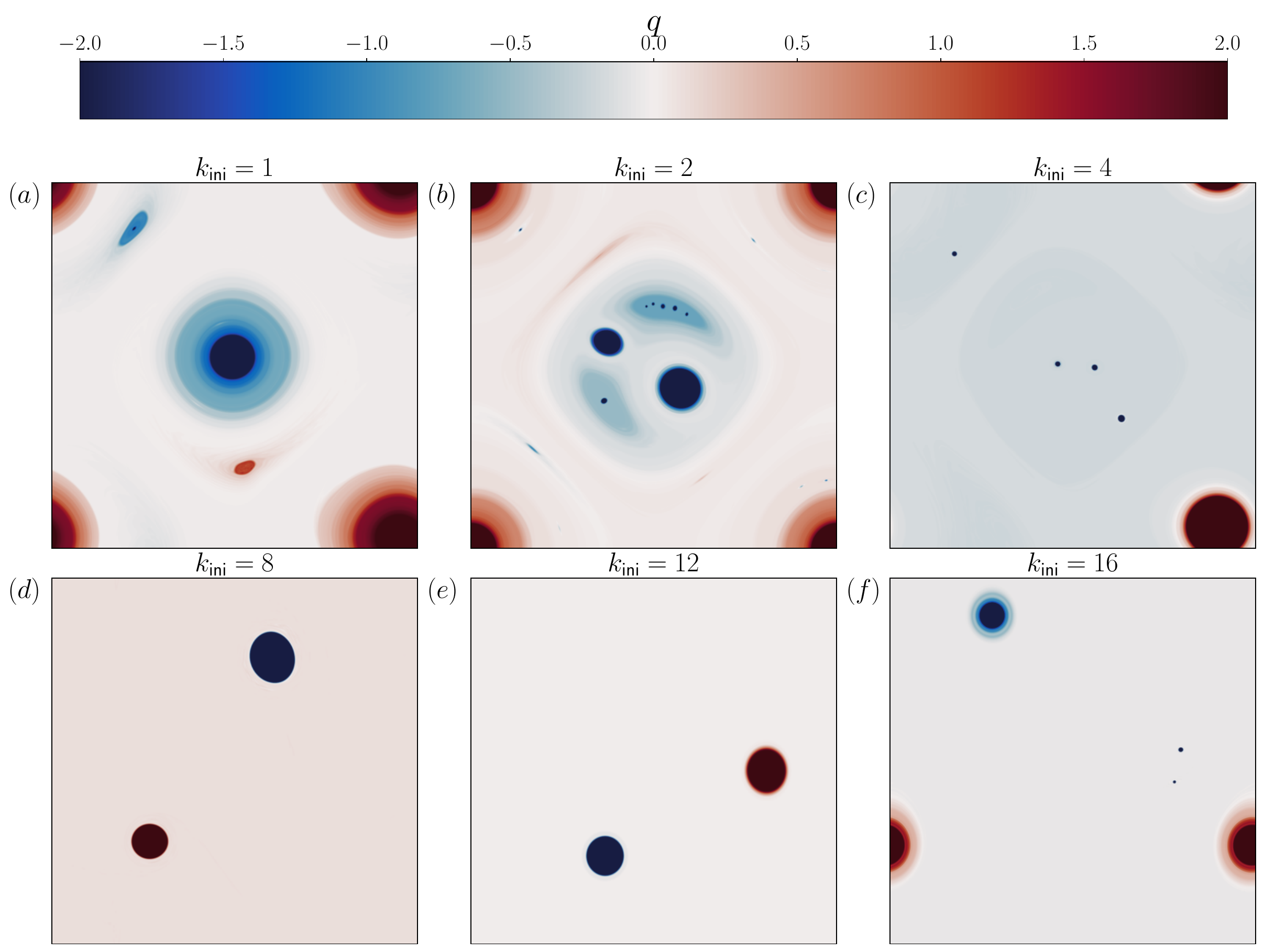}
    \caption{Final states at $t=5000$ for the simulations with the high energy $E=2E_\#$ and initial wavenumbers $k_{\text{ini}}=1-16$.}
    \label{fig:high_energy_results_across_scales}
\end{figure}

In Fig.~\ref{fig:high_energy_results_across_scales}, we present the final snapshots of simulations at the high energy level $E = 2E_\#$ for initial wavenumbers $k_{\text{ini}} = 1$--$16$ at time $t = 5000$.
When the initial condition is at the domain scale ($k_{\text{ini}} = 1$), the final state displays a vortex–topography anti-correlation opposite to that observed in the low-energy regime: an anticyclone is trapped over the central bump, whereas a cyclone aligns with the corner depressions.
As the initial wavenumber increases, this anti-correlation rapidly weakens. The vortices become increasingly energetic and, eventually, freely wander throughout the domain.

\begin{figure}
    \centering
    \includegraphics[width=1\linewidth]{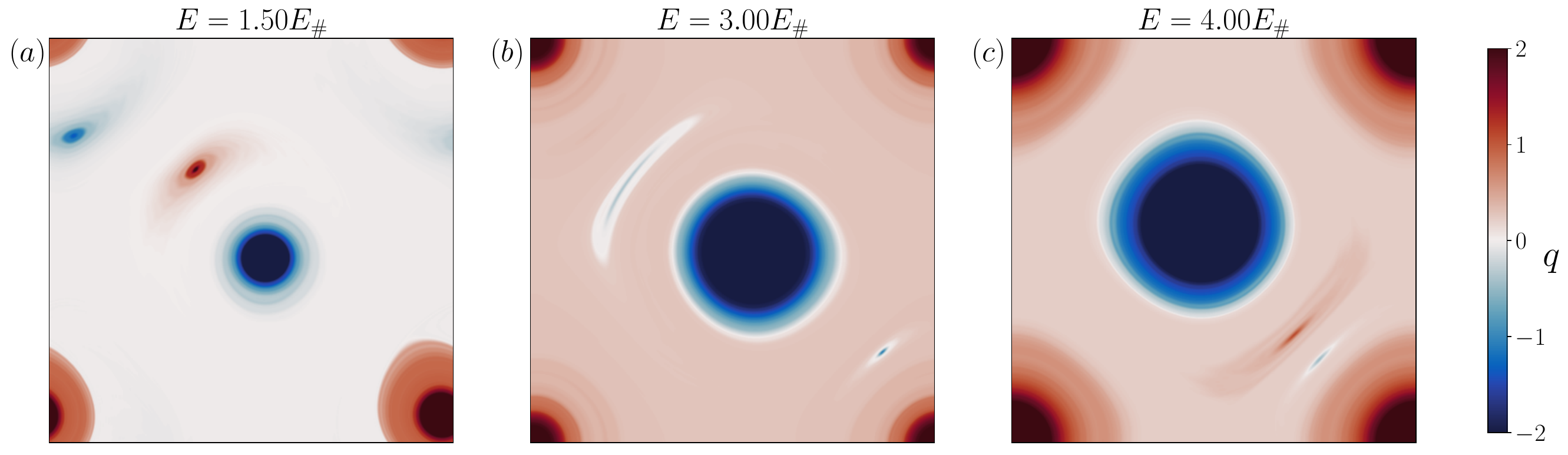}
    \caption{Final states at $t=10000$ for the simulations at the initial wavenumber $k_{\text{ini}}=1$ and three high energy levels of $1.5E_\#$, $3E_\#$ and $4E_\#$.}
    \label{fig:high_energy_results_at_other_energy_levels}
\end{figure}

To further assess the robustness of the vortex–topography anti-correlation at high energy, we conduct additional simulations at other high energy levels while fixing the initial wavenumber at $k_{\text{ini}} = 1$.
The corresponding final states at $t = 10000$ are shown in Fig.~\ref{fig:high_energy_results_at_other_energy_levels} for three energy levels: $1.5E_\#$, $3E_\#$, and $4E_\#$.
To reduce computational cost, these simulations are performed at a resolution of $512$, rather than that of $1024$ used in the cases discussed above, since the initial condition is characterized by a domain-scale wave length.
In all cases, the vortex–topography anti-correlation persists.
These results confirm that, when the initial flow is at the domain scale, a vortex–topography anti-correlation opposite to that found at low energy can robustly emerge at high energy.

\bibliography{final_version}

\end{document}